\documentclass[10pt]{iopart}
\usepackage{graphicx}
\usepackage{amssymb}
\newcommand{\be}{\begin{equation}}
\newcommand{\ee}{\end{equation}}
\newcommand{\bea}{\begin{eqnarray}}
\newcommand{\eea}{\end{eqnarray}}

\def\C{{\cal C}}
\def\V{{\cal V}}
\def\R{{\cal R}}
\def\Z{{\mathbb Z}}
\def\bi{\mathbb{I}}

\def\la{\langle}
\def\ra{\rangle}
\def\d{\partial}
\def\db{\bar\partial}
\def\t{\theta}
\def\tb{\tilde\theta}

\def\mdn{\mu^{D,N}}
\def\mnd{\mu^{N,D}}

\def\conf{
\begin{picture}(0,15)(30,2)
\put(0,0){\circle{10}}
\put(0,0){\makebox(0,0)[c]{\scriptsize 2}}
\put(5,0){\line(1,0){5}}
\put(15,0){\circle{10}}
\put(15,0){\makebox(0,0)[c]{\scriptsize 3}}
\put(20,0){\line(1,0){5}}
\put(30,0){\circle{10}}
\put(30,0){\makebox(0,0)[c]{\scriptsize 2}}
\put(15,5){\line(0,1){5}}
\put(15,15){\circle{10}}
\put(15,15){\makebox(0,0)[c]{\scriptsize 1}}
\put(30,15){\circle{10}}
\put(30,15){\makebox(0,0)[c]{\scriptsize 2}}
\put(-7,0){\line(1,0){2}}
\put(35,0){\line(1,0){2}}
\put(10,15){\line(-1,0){2}}
\put(20,15){\line(1,0){5}}
\put(35,15){\line(1,0){2}}
\put(0,5){\line(0,1){2}}
\put(15,20){\line(0,1){2}}
\put(30,5){\line(0,1){5}}
\put(30,20){\line(0,1){2}}
\put(0,-5){\line(0,-1){2}}
\put(15,-5){\line(0,-1){2}}
\put(30,-5){\line(0,-1){2}}
\end{picture}}

\def\conff{
\begin{picture}(0,15)(30,2)
\put(0,0){\circle{10}}
\put(0,0){\makebox(0,0)[c]{\scriptsize 2}}
\put(5,0){\line(1,0){5}}
\put(15,0){\circle{10}}
\put(15,0){\makebox(0,0)[c]{\scriptsize 3}}
\put(20,0){\line(1,0){5}}
\put(30,0){\circle{10}}
\put(30,0){\makebox(0,0)[c]{\scriptsize 2}}
\put(15,5){\line(0,1){5}}
\put(15,15){\circle{10}}
\put(15,15){\makebox(0,0)[c]{\scriptsize 2}}
\put(30,15){\circle{10}}
\put(30,15){\makebox(0,0)[c]{\scriptsize 1}}
\put(-7,0){\line(1,0){2}}
\put(35,0){\line(1,0){2}}
\put(10,15){\line(-1,0){2}}
\put(20,15){\line(1,0){5}}
\put(35,15){\line(1,0){2}}
\put(0,5){\line(0,1){2}}
\put(15,20){\line(0,1){2}}
\put(30,5){\line(0,1){5}}
\put(30,20){\line(0,1){2}}
\put(0,-5){\line(0,-1){2}}
\put(15,-5){\line(0,-1){2}}
\put(30,-5){\line(0,-1){2}}
\end{picture}}

\def\confff{
\begin{picture}(0,15)(30,2)
\put(0,0){\circle{10}}
\put(0,0){\makebox(0,0)[c]{\scriptsize 2}}
\put(5,0){\line(1,0){5}}
\put(15,0){\circle{10}}
\put(15,0){\makebox(0,0)[c]{\scriptsize 3}}
\put(20,0){\line(1,0){5}}
\put(30,0){\circle{10}}
\put(30,0){\makebox(0,0)[c]{\scriptsize 1}}
\put(15,5){\line(0,1){5}}
\put(15,15){\circle{10}}
\put(15,15){\makebox(0,0)[c]{\scriptsize 2}}
\put(30,15){\circle{10}}
\put(30,15){\makebox(0,0)[c]{\scriptsize 2}}
\put(-7,0){\line(1,0){2}}
\put(35,0){\line(1,0){2}}
\put(10,15){\line(-1,0){2}}
\put(20,15){\line(1,0){5}}
\put(35,15){\line(1,0){2}}
\put(0,5){\line(0,1){2}}
\put(15,20){\line(0,1){2}}
\put(30,5){\line(0,1){5}}
\put(30,20){\line(0,1){2}}
\put(0,-5){\line(0,-1){2}}
\put(15,-5){\line(0,-1){2}}
\put(30,-5){\line(0,-1){2}}
\end{picture}}

\def\congf{
\begin{picture}(0,15)(30,2)
\put(0,0){\circle{10}}
\put(0,0){\makebox(0,0)[c]{\scriptsize 2}}
\put(5,0){\line(1,0){5}}
\put(15,0){\circle{10}}
\put(15,0){\makebox(0,0)[c]{\scriptsize 3}}
\put(20,0){\line(1,0){5}}
\put(30,0){\circle{10}}
\put(30,0){\makebox(0,0)[c]{\scriptsize 2}}
\put(15,5){\line(0,1){5}}
\put(15,15){\circle{10}}
\put(15,15){\makebox(0,0)[c]{\scriptsize 1}}
\put(30,15){\circle{10}}
\put(30,15){\makebox(0,0)[c]{\scriptsize 3}}
\put(45,15){\circle{10}}
\put(45,15){\makebox(0,0)[c]{\scriptsize 1}}
\put(-7,0){\line(1,0){2}}
\put(35,0){\line(1,0){2}}
\put(10,15){\line(-1,0){2}}
\put(20,15){\line(1,0){5}}
\put(35,15){\line(1,0){5}}
\put(50,15){\line(1,0){2}}
\put(0,5){\line(0,1){2}}
\put(15,20){\line(0,1){2}}
\put(30,5){\line(0,1){5}}
\put(30,20){\line(0,1){2}}
\put(45,20){\line(0,1){2}}
\put(45,10){\line(0,-1){2}}
\put(0,-5){\line(0,-1){2}}
\put(15,-5){\line(0,-1){2}}
\put(30,-5){\line(0,-1){2}}
\end{picture}}

\def\congff{
\begin{picture}(0,15)(30,2)
\put(0,0){\circle{10}}
\put(0,0){\makebox(0,0)[c]{\scriptsize 2}}
\put(5,0){\line(1,0){5}}
\put(15,0){\circle{10}}
\put(15,0){\makebox(0,0)[c]{\scriptsize 3}}
\put(20,0){\line(1,0){5}}
\put(30,0){\circle{10}}
\put(30,0){\makebox(0,0)[c]{\scriptsize 2}}
\put(15,5){\line(0,1){5}}
\put(15,15){\circle{10}}
\put(15,15){\makebox(0,0)[c]{\scriptsize 2}}
\put(30,15){\circle{10}}
\put(30,15){\makebox(0,0)[c]{\scriptsize 2}}
\put(45,15){\circle{10}}
\put(45,15){\makebox(0,0)[c]{\scriptsize 1}}
\put(-7,0){\line(1,0){2}}
\put(35,0){\line(1,0){2}}
\put(10,15){\line(-1,0){2}}
\put(20,15){\line(1,0){5}}
\put(35,15){\line(1,0){5}}
\put(50,15){\line(1,0){2}}
\put(0,5){\line(0,1){2}}
\put(15,20){\line(0,1){2}}
\put(30,5){\line(0,1){5}}
\put(30,20){\line(0,1){2}}
\put(45,20){\line(0,1){2}}
\put(45,10){\line(0,-1){2}}
\put(0,-5){\line(0,-1){2}}
\put(15,-5){\line(0,-1){2}}
\put(30,-5){\line(0,-1){2}}
\end{picture}}

\def\congfff{
\begin{picture}(0,15)(30,2)
\put(0,0){\circle{10}}
\put(0,0){\makebox(0,0)[c]{\scriptsize 2}}
\put(5,0){\line(1,0){5}}
\put(15,0){\circle{10}}
\put(15,0){\makebox(0,0)[c]{\scriptsize 3}}
\put(20,0){\line(1,0){5}}
\put(30,0){\circle{10}}
\put(30,0){\makebox(0,0)[c]{\scriptsize 1}}
\put(15,5){\line(0,1){5}}
\put(15,15){\circle{10}}
\put(15,15){\makebox(0,0)[c]{\scriptsize 2}}
\put(30,15){\circle{10}}
\put(30,15){\makebox(0,0)[c]{\scriptsize 3}}
\put(45,15){\circle{10}}
\put(45,15){\makebox(0,0)[c]{\scriptsize 1}}
\put(-7,0){\line(1,0){2}}
\put(35,0){\line(1,0){2}}
\put(10,15){\line(-1,0){2}}
\put(20,15){\line(1,0){5}}
\put(35,15){\line(1,0){5}}
\put(50,15){\line(1,0){2}}
\put(0,5){\line(0,1){2}}
\put(15,20){\line(0,1){2}}
\put(30,5){\line(0,1){5}}
\put(30,20){\line(0,1){2}}
\put(45,20){\line(0,1){2}}
\put(45,10){\line(0,-1){2}}
\put(0,-5){\line(0,-1){2}}
\put(15,-5){\line(0,-1){2}}
\put(30,-5){\line(0,-1){2}}
\end{picture}}

\usepackage[colorlinks=false]{hyperref}
\begin{document}
\title{Pre-logarithmic and logarithmic fields in a sandpile model}
\author{Geoffroy Piroux and Philippe Ruelle\footnote[1]{Chercheur qualifi\'e
FNRS}}
\address{Universit\'e catholique de Louvain\\
Institut de Physique Th\'eorique\\
B--1348 \hskip 0.5truecm Louvain-la-Neuve, Belgium
}
\date{\today}
\begin{abstract}
We consider the unoriented two--dimensional Abelian sandpile model on the
half--plane with open and closed boundary conditions, and relate it to the
boundary logarithmic conformal field theory with central charge $c=-2$. Building on
previous results, we first perform a complementary lattice analysis of the operator
effecting the change of boundary condition between open and closed, which
confirms that this operator is a weight $-1/8$ boundary primary field, whose
fusion agrees with lattice calculations. We then consider the operators
corresponding to the unit height variable and to a mass insertion at an isolated site of
the upper half plane and compute their one--point functions in presence of a boundary
containing the two kinds of boundary conditions. We show that the scaling limit of the
mass insertion operator is a weight zero logarithmic field. 
\end{abstract}



\section{Introduction}

Since about ten years, logarithmic conformal field theories (LCFT) have
been the focus of increasing attention. After the first systematic study
conducted in \cite{gur}, they have prompted intense work of theoretical
developments. A key issue is the proper understanding of their
representation theory, which is considerably more complex than in the more
usual rational non--logarithmic theories. For these matters, we will refer to
the review articles \cite{flohr,gab}, and the references therein.

One of the reasons for which they have been so much studied is their capability
to describe universality classes of two--dimensional critical phenomena with
unusual behaviours, due to non--local or non--equilibrium features. Examples of
lattice systems falling in these classes include polymers, percolation, disordered
systems, spanning trees and sandpile models. 

Sandpile models have been defined by Bak, Tang and Wiesenfeld \cite{btw}, and
proposed as prime examples of open dynamical systems showing generic critical
behaviour (so--called self--organized critical systems). The specific sandpile
model we examine here is the two--dimensional, unoriented Abelian sandpile model
(ASM). It is the model originally considered in \cite{btw}, and remains one of
the simplest but most challenging models. 

It is not clear yet whether the ASM can be fully described by an LCFT, but it is
perhaps the model where this can be most clearly tested, and where the LCFT 
predictions can be most easily and most completely compared with lattice
results, making it a sort of Ising model for logarithmic CFTs. In addition the
conformal fields should have a clean identification in terms of lattice
observables. We believe that it is worth pushing the correspondence in a concrete
and totally explicit model in order to gain  intuition for the somewhat exotic
features of LCFTs. The intricate structure of their Virasoro (or extended
algebras) representations have direct consequences on virtually every aspect of
LCFTs. In particular, the construction of boundary states and the interpretation
of the possible boundary conditions is an important issue to which a lattice
point of view can greatly contribute. 

We will specifically focus on certain aspects of the ASM defined on the upper half plane
(UHP). We will start by extending the analysis of \cite{r} as regards the operator
transforming an open boundary condition into a closed one, or vice--versa. This
operator converges in the scaling limit to an $h=-1/8$ pre--logarithmic field $\mu$,
and by looking at its 4--point function, we will conclude that the two channels of its
boundary fusion,
\be
\V_{-1/8} \times \V_{-1/8} = \V_0^{\rm op} + \R_0^{\rm cl},
\label{mumu}
\ee
must be kept separate, in contradistinction to the bulk fusion. The representations $\V_0$
and $\R_0$ respectively contain those fields which live on an open (Dirichlet) or a closed
(Neumann) boundary. In particular, both contain the identity field, but with different
properties since $\la \bi \ra_D = 1$ while $\la \bi \ra_N = 0$.

Our purpose in this paper is to consider various boundary--bulk
correlations for ASM observables and to compare them with pure LCFT calculations. We 
focus in this article on local observables only, defering examples of non--local ones to
future works. An observable to be discussed below
is the height one variable, already well studied in the literature, and identified in the bulk
with a weight 2 primary field. The other is the insertion of a unit mass
(dissipation) at an isolated site in the bulk of the UHP, which will be shown to
correspond to a logarithmic field $\omega$ of dimension 0, the partner field of
the identity. By using the boundary condition changing field $\mu$, we compute
their 1--point functions on the UHP with mixed boundary conditions, open and
closed, on various stretches of the boundary, and compare them with (numerical)
lattice results. For the logarithmic field $\omega$, we will use the lattice
results to compute its bulk and boundary operator product expansions (OPEs) and
its expansion in terms of boundary fields. In particular, the usual interpretation of the
latter suggests that the chiral fusion $\R_0 \times \R_0$ should contain a channel built
on $\V_0$, 
\be
\R_0 \times \R_0 = \V_0^{\rm op} + \R_0^{\rm cl} + \ldots,
\label{omom}
\ee
and that these two channels should again be kept separate, for the same reason as in
(\ref{mumu}). The dots stand for the other representations which might appear in the
abstract chiral fusion, like the representation $\R_{1}$ \cite{gk1}, and which
presumably contain the fields living on other types of boundary than open or
closed.

The article is organized as follows. In Section 2, we will give a short description of
the Abelian sandpile model and its basic features. We will also recall the basics
of the $c=-2$ LCFT that is relevant here, and give a summary of the present status of the
correspondence ASM/LCFT. 

In Section 3, we investigate the lattice correlations of the boundary condition changing
operator $\mu$ and its fusion (\ref{mumu}). Section 4 considers the unit height
variable in presence of a boundary which contains both open and closed sites, on
the lattice and in the LCFT framework. Section 5 and 6 proceed with a similar
analysis for the insertion of mass at an isolated site, and determine
all relevant OPEs and fusions.


\section{Background}

We start by briefly recalling that model. A more complete account, on this model
and on other self--organized systems, can be found in the review articles
\cite{ip,dhar}.

The model is first defined at finite volume, say on a finite portion $L
\times M$ of the square lattice. At each site $i$ is attached an integer--valued
random variable $h_i$ which can be thought of as the height of the sandpile (the
number of sand grains) at $i$. A configuration $\C$ of the sandpile is a set of
values $\{h_i\}$. A site which has a height $h_i$ in $\{1,2,3,4\}$ is
called stable, and a configuration is stable if all sites are stable. A discrete
time dynamics on stable configurations is then defined as follows.

As a first step, one grain of sand is dropped on a random site $i$ of the current
configuration $\C_t$, producing a new configuration $\C'_t$ which may not be
stable. If $\C'_t$ is stable, we simply set $\C_{t+1}=\C'_t$. If $\C'_t$ is not stable (the
new $h_i$ is equal to 5), it relaxes to a stable configuration $\C_{t+1}$ by letting all
unstable sites topple: a site with height $h_i \geq 5$ loses 4 grains of sand, of which
each of its neighbours receives 1. Relaxation stops when no unstable site remains; the
corresponding stable configuration is $\C_{t+1}$. 

When an unstable site $i$ topples, the updating of all sites will be
written as $h_j \rightarrow h_j-\Delta_{ij}$ for all sites $j$, in terms of
a toppling matrix $\Delta$. If the toppling rule described above applies to
all sites, bulk and boundary, the toppling matrix is the discrete Laplacian with
open boundary conditions, that is, $\Delta_{ij} = 4$ for $j=i$, and $\Delta_{ij} =
-1$ if $j$ is a nearest--neighbour site of $i$. With these toppling rules, the
boundary sites are dissipative: when a boundary site topples, one (or two in case
of a corner site) sand grain leaves the system. This defines the sandpile model
with open boundary condition on all four boundaries.

Closed boundary condition is the other natural boundary condition that we may
impose. Closed boundary sites, when they topple, lose as many sand grains
as their number of neighbours, that is, 3 on a boundary and 2 at a corner, so that
no sand leaves the system. As a consequence, the height variable of a closed
boundary site takes only three values (two at a corner), conventionally chosen to
be 1, 2 and 3. The rows of the toppling matrix labeled by closed boundary sites
are given by $\Delta_{ij} = 3$ for $j=i$, and $\Delta_{ij} = -1$ if $j$ is a
nearest neighbour of $i$. 

Thus closed boundary sites and bulk sites are conservative (diagonal entries of
$\Delta$ equal the coordination numbers so that $\sum_{j} \Delta_{ij} = 0$), whereas    
open boundary sites are dissipative.
One may also make bulk sites dissipative \cite{mkk} by simply increasing the
corresponding diagonal entries $\Delta_{ii}$ from 4 to $4 + t_i$, in which case
we will say that site $i$ has mass $t_i$. (Because when there is enough
dissipation in the bulk, the sandpile enters an off--critical, massive regime, 
characterized by correlations which decay exponentially, and described by a
massive field theory \cite{mr}, see below.) The height variable
$h_i$ of a site of mass $t_i$ takes the values between 1 and $4+t_i$. In this
sense, an open boundary site is a closed boundary site with mass 1, but one could
also consider boundary sites with larger masses. 

In all generality, a specific model is completely defined by giving the values
of the masses, in terms of which the toppling matrix can be written
\be
\Delta_{i,j} = \cases{
t_i + {\rm coord.\ number\ of\ } i & for $i=j$, \cr
-1 & if $i$ and $j$ are {\rm n.n.},}
\ee
with $t_i \geq 0$. Stable sites have height variables $h_i$ between 1 and
$\Delta_{ii}$, unstable sites have $h_i \geq \Delta_{ii} + 1$. In all cases, the
toppling updating rule $h_j \rightarrow h_j-\Delta_{ij}$ applies. Then the
dynamics described above is well--defined provided that not all sites are
conservative, i.e. $t_i > 0$ for at least one site.

Dhar \cite{dhar1} made a detailed analysis of these models. Under very mild
assumptions, he showed that on a finite lattice $\Lambda$, there is a unique
probability measure $P^*_\Lambda$ on the set of stable configurations, that is
invariant under the dynamics. Moreover, $P^*_\Lambda$ is uniform on its support,
formed by the so--called recurrent configurations. The number of recurrent
configurations, which plays the role of partition function, is equal to the determinant 
of the toppling matrix, $Z_\Lambda = \det \Delta_\Lambda$. One is interested in the
infinite volume limiting measure $P^*$, defined through the thermodynamic limit $|\Lambda|
\to \infty$ of the correlations of the finite volume measures. In this article, we will
consider $\Lambda$ going to the discrete plane $\Z^2$ or half plane $\Z \times \Z_{>}$.

The infinite volume limit will of course depend to some extent on the mass values $t_{i}$.
It is however widely believed that the continuum limit of these measures are $c=-2$
logarithmic conformal field theoretic measures, possibly perturbed.

The $c=-2$ LCFT is the simplest and the most studied of all logarithmic theories.
The theory which is thought to be relevant for the ASMs is the rational theory
defined in \cite{gk2}, equivalently the local theory of free symplectic fermions
\cite{k} defined by the action
\be
S = {1 \over \pi} \int \partial \t \db \tb.
\label{lag}
\ee
It has a $\cal W$--algebra built on three dimension 3 fields, which organizes
the chiral conformal theory in six representations. Four of them,
$\V_{-1/8}$, $\V_0$, $\V_{3/8}$ and $\V_1$, are highest weight irreducible
representations constructed on primary fields, while the other two, $\R_0$
and $\R_1$, are reducible but indecomposable, and contain respectively $\V_0$ and
$\V_1$. Of special interest here are the representation $\V_{-1/8}$, with highest
weight field $\mu$, non--local in the fields $\t, \tb$, and the two
representations $\V_0$ and $\R_0$. $\V_0$ has the identity $\bi$ as primary
field, and $\R_0$ has two dimension 0 fields as groundstates, the identity $\bi$
and its logarithmic partner $\omega \sim \, :\!\t\tb\!:$. The corresponding
boundary LCFT has been discussed by several authors \cite{blcft}, and reviewed in
\cite{kaw,ishi}.

Various lattice correlations in the ASM have been computed in order to probe
the adequacy of the description by a $c=-2$ LCFT. The most natural and simplest 
observables, but by no means the only ones, are the height variables, namely
the random variables given by $\delta(h_i-a)$ for $a=1,2,\ldots,\Delta_{ii}$. In this
regard, the height 1 variable, and those for $a>4$ as the case may be, is much different
from the other three ($a=2,3,4$), and is much easier to handle. 

On the infinite plane $\Lambda = \Z^2$, the power law ($\sim r^{-4}$) of the
correlation of two height 1 variables was established in \cite{md}, for the case
where all bulk masses are set to zero, while its exponential decay was proved in
\cite{tk}, when the bulk masses are all equal and non--zero. This was further
investigated in \cite{mr}, where the scaling limit was directly computed for the
mixed correlations of unit height variables and of an other dozen cluster
variables \footnote{The results of \cite{mr} pertaining to the cluster variables called
$S_{10}$ and $S_{11}$  do not refer to the clusters pictured in Fig. 1 of \cite{mr},  
which are not weakly allowed configurations in the sense defined there. Instead the results
for $S_{10}$ and $S_{11}$ reported in Table I should be divided by 4, and refer in each
case to any of the following three clusters
\setlength{\unitlength}{0.25mm}
$$
S_{10} = \hskip  1truecm \mbox{\conf} \hskip 1.7truecm \mbox{\conff}
\hskip 1.7truecm \mbox{\confff}\;\;\;,
\qquad \quad
S_{11} = \hskip 1truecm \mbox{\congf} \hskip 2truecm \mbox{\congff}
\hskip 2truecm \mbox{\congfff}\quad\;\;\;. \nonumber
$$
Also in Eq. (6.2), there is a missing factor $P(1)^{3}$, the minus sign in front of 
$M^{6}/16$ must be suppressed, and there is a missing ``c.c.'' in the third parenthesis.
There are two misprints in (4.8): the denominator of $M^{6}$ should be 8 and not 4, and
the two terms multiplied by $\sqrt{2}/\pi^{2}$ should be separated by a $+$ and  not
a $-$. P.R. thanks Frank Redig, Shahin Rouhani and Monwhea Jeng for discussions about 
the clusters $S_{10}$ and $S_{11}$, and Monwhea Jeng for pointing out the misprints.},  
for all bulk masses equal (to zero or not zero). The scaling limit of the unit height
variable, among others, was identified with a
dimension 2 field made up of simple combinations of $\omega$ and of derivatives of
$\t,\tb$, for the Lagrangian (\ref{lag}) perturbed by a mass term $m^2 \int \omega$.    
These identifications have been recently confirmed by the calculation of all multipoint
correlators, and extended to other local observables \cite{jeng3}. The unit height
variable in presence of an infinite line of massive sites, crossing the whole plane, was
considered in \cite{jeng1}.

A number of calculations have been done on the upper half plane $\Lambda = \Z
\times \Z_{>}$ (or equivalently on an infinite strip). The bulk 1--point function of 
the unit height variable, in presence of an open or a closed boundary, has been
worked out in \cite{bip}, and is consistent with the field identification
mentioned above. The boundary 2--point correlators for all pairs of height
variables have been computed in \cite{iv} again for both an open and a closed
boundary. More recently boundary 3--point correlators were obtained in
\cite{jeng2}, and independently in \cite{pr}, where the same computations were
carried out in the massive regime (all bulk sites have equal mass). Reference
\cite{jeng2} also identifies the insertion of a unit mass (or dissipation) at an
isolated point of the boundary. These results are all consistent
with an LCFT interpretation, for a specific identification of all boundary height
variables in terms of $\t,\tb$, both in the massless and in the massive regime.

All together, these results provide a lattice realization for some of the fields
in the representation $\R_0$, in terms of height variables. A lattice
interpretation for the (chiral) primary field $\mu$ of the representation
$\V_{-1/8}$ was given in \cite{r} as the boundary condition changing operator
between an open and a closed boundary condition. More on this identification will be
given in the next section. 

In what follows we reconsider the height 1 variable and we discuss the insertion
of dissipation at an isolated site in the bulk of the half plane, when both the open and
the closed boundary conditions are imposed on different segments of the boundary. This
allows us to discuss several different issues at the same time: the boundary fusion
of the field $\mu$, mixed boundary/bulk correlators and the boundary and bulk fusion of
the field $\omega$. From now on, we use the conventional model, namely all sites are
conservative except: (i) those on an open boundary which lose one sand grain upon 
toppling, and (ii) the few bulk sites which receive a unit mass. 


\section{The boundary condition changing field}

As explained in the previous section, open/Dirichlet and closed/Neumann are natural 
boundary conditions in the sandpile model, and to our knowledge, the
only known ones. According to general principles of boundary conformal field
theory \cite{cardy}, the change from one boundary condition to the other one is 
implemented by the insertion of a boundary condition changing field, which is also the
ground state of the cylinder Hilbert space with the two boundary conditions on the two
edges. That chiral field $\mu$ was determined in \cite{r}, and shown to be a
primary field of conformal weight $-1/8$. We first recall the analysis which
established this result, and then extend it in order to discuss its boundary
fusion. 

Closing sites on an open boundary or opening sites on a closed boundary changes
the number of recurrent configurations of the sandpile model, and thus the
partition function. Let us denote by $Z_D(I)$ and $Z_N(I)$ the partition functions
of the sandpile model on the UHP with open resp. closed boundary condition all
along the real axis, except on the interval $I$ where the sites are closed
resp. open. For $I=\emptyset$, the boundary is either all open or all closed.

The effect of opening or closing sites on $I$ can be measured from the fraction by
which the number of recurrent configurations increases or decreases, i.e. from the
ratio of partition functions $Z_D(I)/Z_D(\emptyset)$ and $Z_N(I)/Z_N(\emptyset)$.
These ratios correspond to the expectation values of the closing or the opening of the
sites of $I$, and should be given in the scaling limit by the 2--point function of the
two $\mu$ fields located at the endpoints of $I$. For what follows, it is worth
recalling some details on the way these quantities are actually computed.

The partition function of a sandpile model is equal to the determinant of its
toppling matrix. Because opening or closing sites changes by 1 the relevant
diagonal entries in that matrix, the ratios of partition functions are given by
\bea
{Z_D(I) \over Z_D(\emptyset)} &=& {\det{[\Delta_{\rm op} - B_I]} \over
\det{\Delta_{\rm op}}} = \det{[\bi - \Delta_{\rm op}^{-1} \, B_I]} = 
\det(\bi - \Delta_{\rm op}^{-1})_{i,j \in I}\,,
\label{first}\\ 
{Z_N(I) \over Z_N(\emptyset)} &=& {\det{[\Delta_{\rm cl} + B_I]} \over
\det{\Delta_{\rm cl}}} = \det{[\bi + \Delta_{\rm cl}^{-1} \, B_I]} =
\det(\bi + \Delta_{\rm cl}^{-1})_{i,j \in I}\,.
\label{second}
\eea
In these expressions, $\Delta_{\rm op}$ and $\Delta_{\rm cl}$ are the usual
discrete Laplacians on the UHP with either open or closed boundary condition on
the real axis, and $(B_I)_{i,j} = \delta_{i,j} \, \delta(i \in I)$ is a defect
matrix which is used to insert or to remove a unit mass from the sites of $I$. So
$\Delta_{\rm op} - B_I$ is the Laplacian on the UHP with open boundary condition
except on the interval $I$ which is closed, whereas $\Delta_{\rm cl} + B_I$ is the
Laplacian for a closed boundary condition except on $I$ which is open. 

The first ratio (\ref{first}) is simpler. The determinant
$\det(\bi - \Delta_{\rm op}^{-1})_{i,j \in I}$ has finite rank and is finite
since the entries of $\Delta_{\rm op}^{-1}$ are finite. Therefore, although the
two partition functions are infinite, their ratio is a finite number. It is not
difficult to see that this determinant has the Toeplitz form, and can be
evaluated asymptotically when its rank is large, by using a generalization of the
classical Szeg\"o theorem. For an interval $I = [1,n]$, one finds, after
subtracting a non--universal term related to the change of boundary entropy, a
ratio $Z_D(I)/Z_D(\emptyset) \sim A \, n^{1/4}$ for large $n$, for $A
= 1.18894$. As this should be equal to a 2--point correlator, one finds $\la \mdn
(0) \mnd(n) \ra = A \, n^{1/4}$, where the constant $A$ is a structure constant
of the $\mu$ field.

The calculation of the ratio of partition functions for the converse situation
is in principle similar but brings a little though crucial difference. The ratio
is again given by a finite rank determinant $\det(\bi + \Delta_{\rm
cl}^{-1})_{i,j \in I}$, but unlike the previous case, it is infinite because the
entries of $\Delta_{\rm cl}^{-1}$ are infinite (see the Appendix). As the entries of the
matrix $\bi + \Delta_{\rm cl}^{-1}$ all contain the same singular
term $2\Delta^{-1}_{\rm plane}(0;0)$, its determinant has the form 
$2\Delta^{-1}_{\rm plane}(0;0) f(n) + g(n)$. The regularized ratio 
$Z_N(I)/Z_N(\emptyset)$ is defined to be the function $f(n)$, and can be seen as
the partition function $Z_N(I)$ normalized by $Z_N(\{0\}) = 1 + \Delta_{\rm
cl}^{-1}(0;0) = 3/4 + 2\Delta^{-1}_{\rm plane}(0;0)$, the partition
function for a closed boundary with a single boundary site open. 

The regularized determinant itself is given by a determinant. Set $M_{ij} = (\bi +
\Delta_{\rm cl}^{-1})_{ij \in I}$. We define a new matrix $M'$ from $M$ by first 
subtracting the first row of $M$ from the other rows, and then subtracting the first
column of the new matrix from the other columns. The resulting matrix $M'$ has by 
construction the same determinant as $M$, but has
finite entries, except $M'_{11}$ which is the only one to contain the infinite
term 
$2\Delta^{-1}_{\rm plane}(0;0)$,
\bea
M'_{11} &=& M_{11}\,, \qquad M'_{1j} = M_{1j} - M_{11},
\quad (j>1) \qquad M'_{i1} = M_{i1} - M_{11},\quad (i>1) \\
M'_{ij} &=& M_{ij} - M_{1j} - M_{i1} + M_{11}. \quad (i,j>1)
\eea
Thus the coefficient of $2\Delta^{-1}_{\rm plane}(0;0)$ in $\det M$ is equal to the
(1,1) minor of $M'$,
\be
f(n) \equiv \det(\bi + \Delta_{\rm cl}^{-1})_{\rm reg} = \det (M')_{i,j > 1}.
\label{reg}
\ee

The regularized ratio $Z_N(I)/Z_N(\emptyset)$ can be computed exactly when    
the size of $I=[1,n]$ gets large. One finds exactly the same result as in the open case,
namely that $Z_N(I)/Z_N(\emptyset) \sim A \, n^{1/4}$ for large $n$, with the same 
value of the constant $A$. As expected, it implies the same 2--point function $\la \mnd (0)
\mdn(n) \ra = A \, n^{1/4}$ as before.

These results suggest that the field changing a boundary condition from open to
closed or vice--versa is a chiral primary field of weight $-1/8$. The way they
fuse on the boundary depend on their relative positions. The OPE $\mdn(0)
\mnd(z)$ must close on fields that live on a $D$ boundary, i.e. on fields in the
representation ${\cal V}_0$, while the other OPE $\mnd(0) \mdn(z)$ should
only contain fields that live on an $N$ boundary, and therefore taken from the
representation $\R_0$ \cite{r}. Hence one infers that
\bea
\mdn(z) \, \mnd(0) &=& z^{1/4} \, C_{\mu,\mu}^{\bi_D} \, \bi_D(0) +
\ldots \label{fusD}\\
\noalign{\medskip}
\mnd(z) \, \mdn(0) &=& z^{1/4} \, C_{\mu,\mu}^\omega \,  [ \lambda_{N} \,
\bi_N(0) \, \log{z} + \omega_N(0)] + \ldots \label{fusN}
\eea
where subscripts have been added to stress the type of boundary the fields live
on. The coefficient $\lambda_{N}$ specifies the transformation 
\be
\omega_{N}(z) \to \omega_{N}(w) + \lambda_{N} \log{{\rm d}w
\over {\rm d}z}
\ee 
under a chiral conformal transformation $z \to w(z)$. 

For these OPEs to be consistent with the 2--point functions given above, one
should have, for a choice of normalizations, 
\bea
\la \bi_D \ra_{\rm op} =&& 1\,, \qquad C_{\mu,\mu}^{\bi_D}=A\,, \\
\noalign{\medskip}
\la \bi_N \ra_{\rm cl} =&& 0\,, \qquad \la \omega_N \ra_{\rm cl} = 1\,, \qquad
C_{\mu,\mu}^{\omega}=A\,.
\eea
These equations emphasize the fact that $\bi_D \in \V_0$ and $\bi_N \in \R_0$ are
genuinely different fields, and that the two conformal blocks in the $\mu\mu$ OPE
must be kept separate. This is in contrast with the fusion in the bulk, where the
two identities have to be identified, so that the fields in the $\V_0$ channel
are to be considered as a subset of those in the $\R_0$ channel. 

Let us also note that in the lattice ASM interpretation, the limit
$\lim_{z \to 0}\,{1 \over Az^{1/4}} \la \mnd (0) \mdn(z) \ra = 1$ should
correspond to the expectation value of the identity in presence of a closed
boundary condition, and that this is actually obtained by selecting the $\omega$
channel in the OPE. One may see this as the trace left by the regularization used in the
computation of the ratio of partition functions, or equivalently by the single open
site left in the otherwise closed boundary. It fits the picture we are going to give in
the next sections, where the insertion of a unit mass at an isolated closed site
corresponds to the insertion of a field $\omega$ (see also \cite{jeng2}). By applying the
same arguments to a finite (in the continuum) interval $[0,z]$, one could see the opening
of that interval as the insertion of a massive defect line, 
\be
\mnd (0) \mdn(z) \sim \int_0^z \, {\rm d}x \, \omega_N(x).
\ee

Instead of closing or opening sites on a segment, one can do it on two segments,
$I_1 = [z_1,z_2]$ and $I_2=[z_3,z_4]$, a situation that was only briefly
discussed in \cite{r}. In the scaling limit where all distances $|z_i - z_j|$ are
large with finite ratios, the appropriate ratios of partition functions
$Z_D(I_1,I_2)/Z_D(\emptyset)$ and $Z_N(I_1,I_2)/Z_N(\emptyset)$ (regularized) are  
expected to converge to the appropriate 4--point function
\bea
{Z_D(I_1,I_2) \over Z_D(\emptyset)} & \longrightarrow & \la \mdn(z_1) \mnd(z_2)
\mdn(z_3) \mnd(z_4) \ra\,, \label{dii}\\
\noalign{\medskip}
{Z_N(I_1,I_2) \over Z_N(\emptyset)} & \longrightarrow & \la \mnd(z_1) \mdn(z_2)
\mnd(z_3) \mdn(z_4)  \ra \,. \label{nii}
\eea

The irreducible representation $\V_{-1/8}$ being degenerate at level 2, the 4--point
function of $\mu$ satisfies a second order differential equation. Introducing the cross
ratio of the four insertion points $x = {z_{12}z_{34} \over z_{13}z_{24}}$ with
$z_{ij}=z_i-z_j$, one finds the general form \cite{gur}
\be
\la \mu(z_1) \mu(z_2) \mu(z_3) \mu(z_4) \ra = (z_{12} z_{34})^{1/4} \, 
(1-x)^{1/4} \,[\alpha \, K(x) + \beta \, K(1-x)]\,,
\label{4pt}
\ee
in terms of the complete elliptic integral
\be
K(x) = \int_0^{\pi \over 2} \, {{\rm d}t \over \sqrt{1 - x \sin^2{t}}} = 
\cases{\textstyle {\pi \over 2} + \ldots & for $x \sim 0^+$,\cr
\textstyle -{1 \over 2} \log{1-x \over 16} + \ldots & for $x \sim 1^-$.}
\ee
If we assume the ordering $z_1 < z_2 < z_3 < z_4$, the cross ratio
$x$ is real and positive in $[0,1]$.

To verify the asymptotic behaviours (\ref{dii}) and (\ref{nii})
explicitly, the coefficients $\alpha$ and $\beta$ must be fixed in each case. 

In the first case, namely two closed segments in an open boundary, the ratio 
\be
{Z_D(I_1,I_2) \over Z_D(\emptyset)} = 
{\det{[\Delta_{\rm op} - B_{I_1} -  B_{I_2}]} \over \det \Delta_{\rm op}} =  
\det[\bi - \Delta_{\rm op}^{-1}]_{i,j \in I_{1} \cup I_{2}}
\ee
is a determinant of dimension $|I_{1}| + |I_{2}|$. It contains a `disconnected' piece
proportional to the product $[{Z_D(I_1)/Z_D(\emptyset)}] [{Z_D(I_2)/
Z_D(\emptyset)}]$, plus connected contributions that involve off--diagonal entries 
$\Delta_{\rm op}^{-1}(i,j)$ with $i$ in $I_1$ and $j$ in $I_2$, or vice--versa. These 
entries decay like the inverse distance between $i$ and $j$, a distance bounded
below by $|z_{32}|$, implying that the off--diagonal blocks go to 0 when this distance
goes to infinity,
\be
\lim_{z_{32} \to +\infty} {Z_D(I_1,I_2) \over Z_D(\emptyset)} = {Z_D(I_1)
\over Z_D(\emptyset)} \, {Z_D(I_2) \over Z_D(\emptyset)} \longrightarrow \la
\mu(z_1) \mu(z_2) \ra \; \la \mu(z_3) \mu(z_4) \ra = A^2 (z_{12} z_{34})^{1/4}.
\ee

As the limit $z_{32}$ going to $+\infty$ means $x$ going to 0, the previous
constraint fixes unambiguously the constants to $\alpha = {2A^2 \over \pi}$ and
$\beta=0$, yielding
\be
\la \mdn(z_1) \mnd(z_2) \mdn(z_3) \mnd(z_4) \ra = {2 A^2 \over \pi} \,  (z_{12}
z_{34})^{1/4} \, (1-x)^{1/4} \, K(x).
\label{dn}
\ee

If the lengths of $I_1$ and $I_2$ are large enough so that
$Z_D(I_1)/Z_D(\emptyset)$ and $Z_D(I_2)/Z_D(\emptyset)$ can be approximated by
$A \, |z_{12}|^{1/4}$ and $A |z_{34}|^{1/4}$, then the intended verification of
(\ref{dii}) can be restated quite concretely as the statement that the following
convergence holds
\be
{Z_D(I_1,I_2) Z_D(\emptyset) \over Z_D(I_1) Z_D(I_2)} = 
{\det{[\bi - \Delta_{\rm op}^{-1} \, B_{I_1} - \Delta_{\rm op}^{-1} \, B_{I_2}]}
\over \det[\bi - \Delta_{\rm op}^{-1} \, B_{I_1}] \, 
\det[\bi - \Delta_{\rm op}^{-1} \, B_{I_2}]}
\longrightarrow {2 \over \pi} \,  (1-x)^{1/4} \, K(x).
\label{ellop}
\ee

\begin{figure}[h]
\leavevmode
\begin{center}
\mbox{\includegraphics[width=8cm]{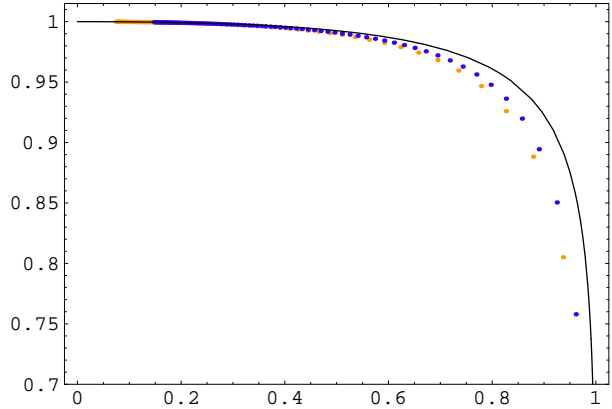}}
\end{center}
\caption{\small Ratio of the ASM partition functions in Eq.(\ref{ellop}) for two
segments of closed sites ($[1,n]$ and $[n+N,2n+N]$) in an open boundary, as
function of the anharmonic ratio $x$. The solid curve is the CFT prediction, the
other two are numerical: $n=30$ (orange) and $n=50$ (blue).}
\end{figure}

Unlike the determinants for one interval, we have not been able to do an exact
asymptotic analysis of the determinant for two intervals. The ratio of
lattice determinants (all well--defined and finite) have been computed
numerically and plotted in Figure 1 as colour dots (details on the numerical aspects of the
calculations are given in the Appendix). We have taken
two segments of equal and fixed length $n$, lying $N$ sites apart. For fixed $n$ (taken
to be 30 and 50), the distance $N=z_{32}$ was varied from 1 to 80, and the
numerical results plotted as a function of $x = ({n \over n+N})^2$, which ranges
between 0 and 1 as $N$ varies. The scaling regime corresponds to large
$n$ and large $N$, and within this setting, is best approached for
large enough values of $N$, that is for values of $x$ not too close to
1. The CFT prediction for the corresponding quantity, namely the
function on the right side of Eq.(\ref{ellop}), is plotted as a solid line.

The agreement is more than satisfactory in the region closed to the scaling
regime, and when $x$ is close to~1, it improves with larger values of $n$.
In particular, this supports the view that the correlator $\la \mdn \mnd \mdn
\mnd \ra$ is regular at $x=0$, and the absence of the conformal block related to
$K(1-x)$. 

The OPEs quoted earlier in this section readily follow from this correlator. When
$z_{12}$ goes to 0, $x$ goes to 0 as well, and the expansion yields
\be
\la \mdn(z_1) \mnd(z_2) \mdn(z_3) \mnd(z_4) \ra = A^2 \, z_{12}^{1/4} \,
z_{34}^{1/4} + \ldots,
\ee
as expected.

If $z_{23}$ goes to 0, then $x$ goes to~1 where the same correlator develops a
logarithmic singularity, and we find
\be
\la \mdn(z_1) \mnd(z_2) \mdn(z_3) \mnd(z_4) \ra = -{A^2 \over \pi} \,
z_{23}^{1/4} \,
\Big[z_{14}^{1/4} \, \log{z_{23}} + z_{14}^{1/4} \, \log{z_{14} \over 16 z_{13}
z_{34}} \Big] + \ldots
\ee
which implies ($\lambda_{N}$ is the value of the off--diagonal entry in the
rank 2 Jordan block)
\be
\lambda_{N} = -{1 \over \pi}.
\label{lN}
\ee

\begin{figure}[htpb]
\leavevmode
\begin{center}
\mbox{\includegraphics[width=8cm]{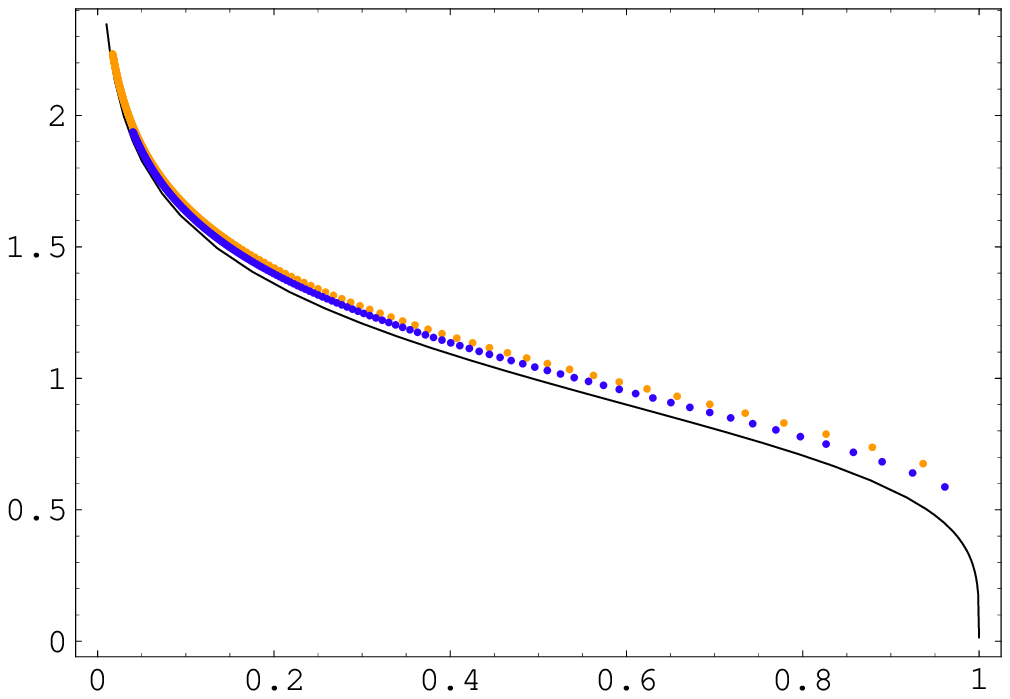}} \hfill
\mbox{\includegraphics[width=8cm]{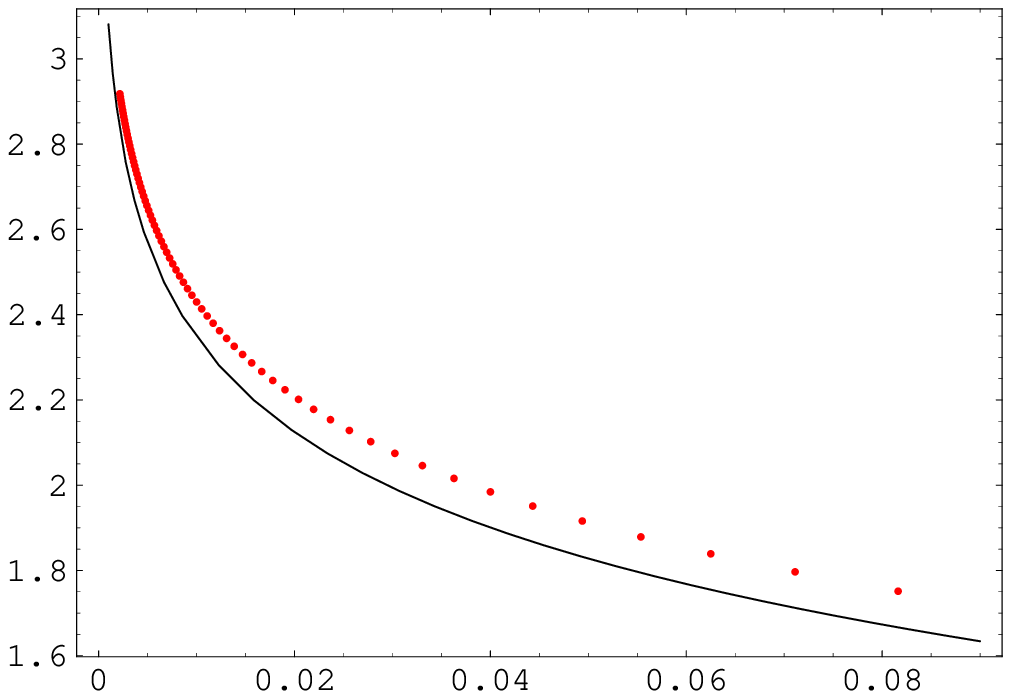}}
\end{center}
\caption{\small Ratio of the ASM partition functions (\ref{ellcl}) for two
segments of open  sites ($[1,n]$ and $[n+N,2n+N]$) in a closed boundary, as
function of the anharmonic ratio $x$. In both figures, the solid curve is the CFT
prediction, given on the right side of Eq.(\ref{ellcl}). The colour dots show the
numerical results: in Fig. 2a, $n=30$ in orange and $n=50$ in blue, where, in both
cases, $N$ is varied from 1 up to 200 sites; in Fig.2b, $n=20$ with $N$ running
between 50 and 410.}
\end{figure}

Let us briefly discuss the second situation, with two open segments in a
closed boundary. The relevant correlator $\la \mnd \mdn \mnd \mdn \ra$ can be
obtained from the previous one by a cyclic permutation of the insertion points,
e.g. the permutation $(1,2,3,4) \to (2,3,4,1)$ (it can also be obtained from an
inversion, a method we will use in Section 4). This changes $x$ into $1-x$, the prefactor
$(z_{12} z_{34})^{1/4} \, (1-x)^{1/4}$ stays invariant, and one readily
obtains
\be
\la \mnd(z_1) \mdn(z_2) \mnd(z_3) \mdn(z_4) \ra = {2 A^2 \over \pi} \, 
(z_{12} z_{34})^{1/4} \, (1-x)^{1/4} \, K(1-x).
\label{nd}
\ee
It makes use of the other conformal block, and as a consequence, is now
logarithmically divergent at $x=0$.

Proceeding as before, one is led to verify the convergence
\be
{Z_N(I_1,I_2) Z_N(\emptyset) \over Z_N(I_1) Z_N(I_2)} 
\longrightarrow {2 \over \pi} \,  (1-x)^{1/4} \, K(1-x).
\label{ellcl}
\ee

The ratio of regularized determinants has been computed numerically, in the
same setting as for the other situation, namely two open segments of length equal
to 30 (orange in Figure 2a) and to 50 (blue). We have included larger distances
$N$ between the two intervals, namely $N$ running from 1 to 200 sites, so that the
variable $x$ could assume smaller values. In order to approach even better the
small $x$ region, where a logarithmic divergence is expected, we have also
considered slightly shorter intervals, of length equal to 20, but separated by
larger distances, up to 410 sites, which allows a minimal value of $x$ equal to
0.0022 (see Figure 2b). 


\section{The unit height variable}

In this section we examine a first instance of a correlator including boundary and
bulk operators. The lattice quantity we consider is the probability $P_1(z)$ that
a certain site $z$ in the UHP has a height variable equal to~1, given a boundary
condition on the real axis mixing both open and closed sites. 

On a finite lattice $\Lambda$, it was shown in \cite{md} that the number of
recurrent configurations with $h_z = 1$ is equal to the total number of recurrent
configurations of a new sandpile model. The new model is defined from the original
one by cutting off the bonds between the site $z$ and any three of its four neighbours,
and by reducing the toppling threshold of $z$ from 4 to~1, and the threshold of the three
neighbours from 4 to 3. Thus the toppling matrix of the new model is equal to
$\Delta_\Lambda^{\rm new} = \Delta_\Lambda - B_z$, where the matrix $B_z$ is
zero everywhere in $\Lambda$ except at $z$ and the three neighbours, where it
reads (first label corresponds to the site $z$)
\be
B_z = \pmatrix{3 & -1 & -1 & -1 \cr
-1 & 1 & 0 & 0 \cr -1 & 0 & 1 & 0 \cr -1 & 0 & 0 & 1}.
\ee

Since $\det \Delta_\Lambda$ is the total number of recurrent configurations in a
sandpile model of toppling matrix $\Delta_\Lambda$, the probability $P_1(z)$
is given by the ratio $\det \Delta_\Lambda^{\rm new}/\det \Delta_\Lambda$. In the
infinite volume limit $\Lambda \to$ UHP, one has
\be
P_1(z) = {\det{[\Delta - B_z]} \over \det \Delta} = \det{[\bi - \Delta^{-1} \,
B_z]},
\ee
a  4--by--4 determinant. Here $\Delta$ is the Laplacian on the UHP subjected to whatever
boundary condition we choose on the boundary. 

The boundary conditions we want to consider are the same as in the previous
section: an open boundary with closed sites on a finite interval $I=[z_1,z_2]$,
or vice--versa, a closed boundary with open sites on $I=[z_1,z_2]$. We define
two probability functions relative to these two situations, $P^{\rm op}_1
(z_1,z_2;z)$ and $P^{\rm cl}_1(z_1,z_2;z)$. Their limit when $\Im z \to +\infty$
yields the bulk probability for any given site to have height 1, $P_1 = P^{\rm
op}_1(z_1,z_2;+i \infty) = P^{\rm cl}_1(z_1,z_2;+i \infty) = 2(\pi - 2)/\pi^3$
\cite{md}.

The lattice calculation of the two probabilities is straightforward. To the two
boundary conditions correspond the Laplacians $\Delta_{\rm op} - B_I$ or 
$\Delta_{\rm cl} + B_I$, where $(B_I)_{i,j} = \delta_{i,j} \, \delta(i \in I)$ is
the matrix used in the previous section, and one obtains
\be
P_1^{\rm op}(z_1,z_2;z) = {\det{[\bi - \Delta_{\rm op}^{-1}\,(B_I + B_z)]}
\over \det{[\bi - \Delta_{\rm op}^{-1}\,B_I]}}\,, \quad 
P_1^{\rm cl}(z_1,z_2;z) = {\det{[\bi + \Delta_{\rm cl}^{-1}\,(B_I - B_z)]} \over
\det{[\bi + \Delta_{\rm cl}^{-1}\,B_I]}}\,.
\label{p1opcl}
\ee
One may take $I=[-R,R]$ centered around the origin, so that the probabilities are
the ratio of a rank $2R+5$ determinant by a rank $2R+1$ determinant. We note that
although the entries of $\Delta_{\rm cl}^{-1}$ contain the divergent piece we
called $2 \Delta^{-1}_{\rm plane}(0;0)$ in Section 3, the ratio defining $P_1^{\rm
cl}(z_1,z_2;z)$ is well--defined and finite. If $I$ is not empty, the two
determinants are proportional to $2 \Delta^{-1}_{\rm plane}(0;0)$ (the proportionality
factor for the denominator, a function of $z_1 - z_2$, goes to $A \, z_{12}^{1/4}$
in the scaling limit), and the ratio is finite. If $I$ is empty, the probability
$P_1^{\rm cl}(z)$ reduces to $\det{[\bi - \Delta_{\rm cl}^{-1}\,B_z]}$, which is
finite because the row and column sums of $B_{z}$ are zero, with the consequence 
that $\Delta_{\rm cl}^{-1}\,B_z$ depends on differences of $\Delta_{\rm cl}^{-1}$ 
entries only, which are well--defined.

Precisely in the extreme case when the interval $I$ is empty, the two probabilities 
$P_1^{\rm op}(z)$  and $P_1^{\rm cl}(z)$ for having a height 1 at site $z$ in front of
an all open or an all closed boundary are given by 4--by--4 determinants. For $z = x + i
y$, the expansion of the two determinants in powers of $y$ yields \cite{bip}
\be
P^{\rm op}_1(z) = P_1 + {P_1 \over 4 y^2} + \ldots\,, \qquad
P^{\rm cl}_1(z) = P_1 - {P_1 \over 4 y^2} + \ldots
\ee

The two functions  $P_1^{\rm op}(-R,R;z)$ and  $P_1^{\rm cl}(-R,R;z)$ have been
computed numerically for various values of $R$ and $z$. The results for $P_1^{\rm
op}(-R,R;z)$ along different lines in the UHP are pictured in Figure 3. 

\begin{figure}[h]
\leavevmode
\begin{center}
\mbox{\includegraphics[width=8cm]{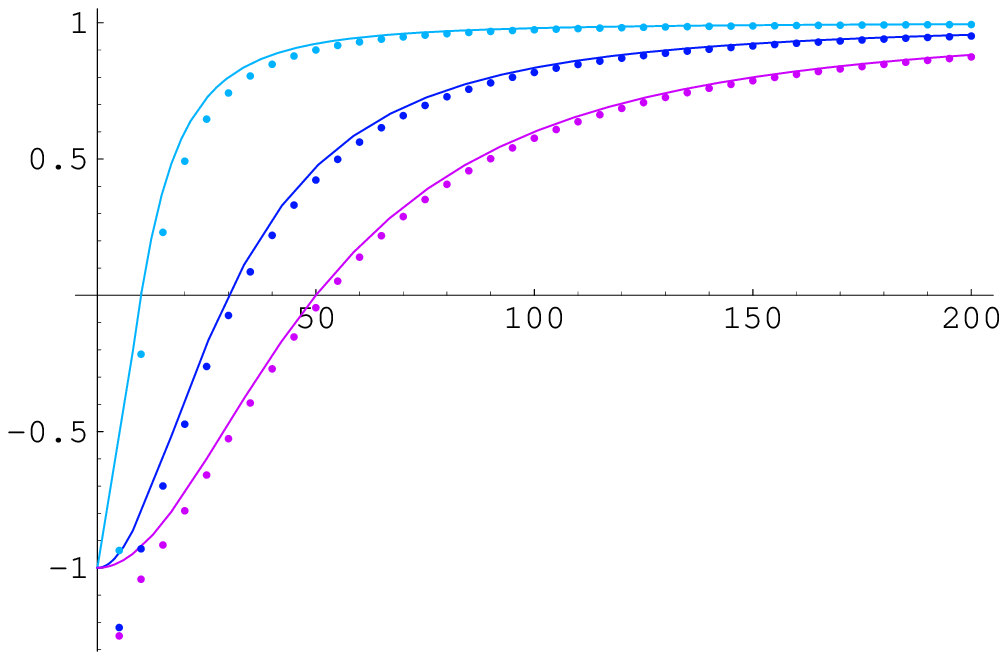}} \hfill
\mbox{\includegraphics[width=8cm]{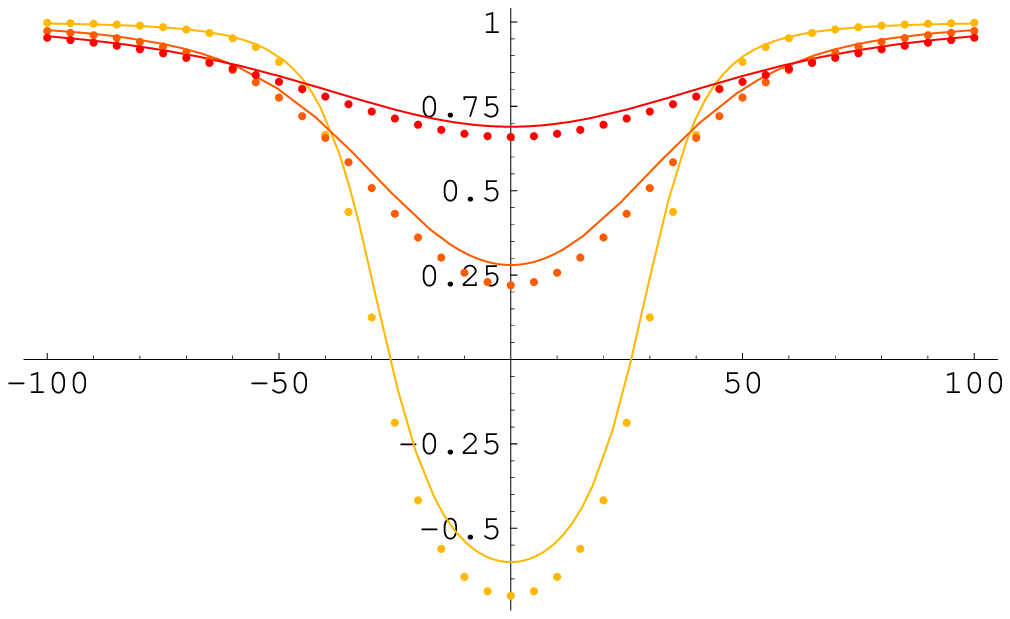}}
\end{center}
\caption{\small Values of $4y^2 \, {P^{\rm op}_1(-R,R;x+i y) \over P_1} - 1 $ as 
function  of $y$ (left) or as function of $x$ (right), either computed numerically (dots) or
from CFT computations, Eq. (\ref{p1}) (solid lines). The curves on the left
correspond to $x=0$ and $5 \leq y \leq 200$, for values of $R$ equal to $R=10$ (light
blue), $R=30$ (blue) and $R=50$ (violet). Those on the right correspond to $R=30$
and $y=15$ (light orange), $y=40$ (orange) and $y=70$ (red), with $-100 \leq x
\leq 100$.}
\end{figure}

If $\phi(z)$ is the scaling field to which the random lattice variable $\delta(h_z-1)
-  P_{1}$ converges in the scaling limit, $P^{\rm op}_1(z_1,z_2;z)$ should be given by the
following correlation
\be
P^{\rm op}_1(z_1,z_2;z) - P_1 = 
{\la \mdn(z_1) \, \mnd(z_2) \, \phi(z) \ra _{\rm UHP} \over \la \mdn(z_1) \, 
\mnd(z_{2})  \ra}\,.
\ee
The 3--point correlation function on the UHP can be rewritten as a 4--point chiral
correlation on the whole plane
\be
P^{\rm op}_1(z_1,z_2;z) - P_1 = {1 \over A z_{12}^{1/4}} \; 
\la \mdn(z_1) \, \mnd(z_2) \, \phi(z) \, \phi(z^{*}) \ra_{\rm chiral,plane}\,.
\ee

In concrete terms, the field $\phi(z)$ is a primary field of weight (1,1), proportional to
$\d \t \db \tb + \d \tb \db \t$ in the $\t,\tb$ model \cite{mr}. The relevant
4--point function is then easy to compute. It satisfies two second order differential
equations: one coming from $\mu$, the other from $\phi$. By combining them, one obtains a
first order differential equation, whose solution yields the general form of the chiral 4--point
correlator ($x = {z_{12}z_{34} \over z_{13}z_{24}}$)
\be
\la \mu(z_1) \, \mu(z_2) \, \phi(z_3) \, \phi(z_4) \ra = \alpha \, {z_{12}^{1/4}
\over z_{34}^2} \, {2-x \over \sqrt{1-x}}.
\ee
The constant $\alpha$ can be fixed by requiring that the expression 
\be
P^{\rm op}_1(z_1,z_2;z) - P_1 = -{\alpha \over 4A \, y^2} \, {2-x \over    
\sqrt{1-x}}.
\label{mmpp}
\ee
reduces to $+P_{1}/4y^{2}$ when $z_{1}$ goes to $z_{2}$.

Clearly $z_{12}$ going to 0 corresponds to $x$ going to 0. However, $z_{12}$ can
go to 0 in two ways: either $z_1$ and $z_2$ meet at a finite point of the real
axis, or they meet at infinity, by going ``around'' the equator of the Riemann
sphere. In both cases, $x$ approaches 0, but from different directions. To see this, we
set $z_1=-R$ and $z_2=R$, and consider $1-x$ for $z_3 = z_4^{*} = z$
\be
1-x = {z_{14} z_{23} \over z_{13} z_{24}} = 
{(R-z)(R+z^{*}) \over (R-z^{*})(R+z)}.
\ee
This expression makes it manifest that $1-x$ has unit modulus and circles 
around 0 when $R$ ranges from 0 to infinity. The expansion of $x$
around $R=0$ and $R=+\infty$,
\be
x = \cases{-{4iyR \over |z|^2} + \ldots & for $R \sim 0$, \cr
{4iy \over R} + \ldots & for $R \sim +\infty$,}
\label{beh}
\ee
shows that $x$ draws out a unit circle centered at 1, travelled anticlockwise from
$-i0$ to $+i0$ as $R$ ranges from 0 to $+\infty$.

One can write $x = 1 - e^{i\theta}$ with $\theta \sim 0^+$ for $R \sim 0^+$,
and $\theta \sim 2\pi^-$ when $R \sim +\infty$, implying that $\lim_{R \to
0} \, \sqrt{1-x} = +1$ and $\lim_{R \to +\infty} \, \sqrt{1-x} = -1$.
Therefore when expanding the correlator (\ref{mmpp}) around $x=0$, one
takes the positive square root $+\sqrt{1-x}$ when $R$ is close to 0, and
the negative root $-\sqrt{1-x}$ when $R$ is close to infinity.

Bearing this in mind, the expansion appropriate to recover the
fully open boundary is around $R=0$, so as to shrink the closed portion to nothing. 
This fixes $\alpha = -{A P_1 \over 2}$, and in turn, yields an explicit expression for the
probability in the scaling limit
\be
P^{\rm op}_1(-R,R;z) - P_1 = -{P_1 \over 4y^2} \, {R^2-|z|^2 \over     
|R^2-z^2|}.
\label{p1}
\ee
This expression is compared with a lattice calculation of the probability in Figure 3. 
The agreement is quite good in view of the fact that the typical distances  
used on the lattice remain modest. 

The same comparison has been made for the converse situation, in which the real axis 
is closed, except for the interval $[z_1,z_2]$ which is open. The corresponding probability
\be
P^{\rm cl}_1(z_1,z_2;z) - P_1 = {1 \over A z_{12}^{1/4}} \; 
\la \mnd(z_1) \, \mdn(z_2) \, \phi(z) \ra_{\rm UHP}\,,
\label{p1cl}
\ee
may be obtained from the previous one by the inversion $z \to -1/z$, which
exchanges the boundary condition around 0 with that around infinity. 
The function $P^{\rm op}_1$ is the ratio of a 4--point and a 2--point function,
which are both invariant under a global conformal transformation, so that the
inversion itself has no effect at all. However, the function has a non--trivial
monodromy around $x=1$, so that the inversion interchanges the two
behaviours (\ref{beh}) around $R=0$ and $R=+\infty$, implying that it must
include a change of sign of all factors $\sqrt{1-x}$. One finds that
\be
P^{\rm cl}_1(-R,R;z) - P_1 = {P_1 \over 4y^2} \, {R^2-|z|^2 \over     
|R^2-z^2|}
\ee
is the opposite of $P^{\rm op}_1(-R,R;z)$. We observe that this probability is
given by the correlator (\ref{p1cl}) through the $\omega$ channel in the
fusion of the two $\mu$'s, since the field $\phi$ does not couple to the identity.
As noted before, this is a remnant of the regularization used to compute
probabilities when the boundary is closed.  

A numerical calculation of this function on the lattice has been carried out for the same
range of parameters as in the previous case. The results were fully consistent with the
predicted change of sign.


\section{Isolated dissipation}

In this section and the next one, we introduce dissipation (or mass) at an
isolated site located in the bulk of the UHP or on a closed boundary. The amount
of dissipation does not really matter, so we consider a minimally dissipative
site, with unit mass. It is convenient to start with the case where a single
boundary condition is imposed along the boundary. The case where two different boundary
conditions coexist is treated in the next section.

>From our discusion in Section 2, introducing dissipation at a given site, located at $z$
say, corresponds to raising the toppling threshold of that site from 4 to 5.
This has the consequences that the height at $z$ takes its values in $\{1,2,3,4,5\}$,
and that each time it topples, one grain of sand is dissipated.  In terms of the
toppling matrix, the introduction of dissipation at $n$ sites $z_{k}$ corresponds
to going from $\Delta$ to $\Delta + D_{z_{1}} + \ldots + D_{z_{n}}$, where $D_z =
\delta_{i,z}\,\delta_{j,z}$ is the defect matrix which makes the site $z$
dissipative (one could say open). 

One can again measure the effect of introducing dissipation by computing
the fraction by which the number of recurrent configurations increases, given
by the ratio of partition functions,
\be
F(z_{1},\ldots,z_{n}) = {\det{[\Delta + D_{z_{1}} + \ldots + D_{z_{n}}]}  \over
\det{\Delta}} = \det[\bi + \Delta^{-1}]_{i,j \in \{z_{1},\ldots,z_{n}\}},
\label{gen}
\ee
and which reduces to the calculation of an $n$--by--$n$ determinant.

This fraction $F(z_{1},\ldots,z_{n})$ is related to the probability that all
sites of the lattice have a height smaller or equal to 4, or to the
probability that one of the sites $z_{i}$ has a height equal to 5, 
\be
F(z_{1},\ldots,z_{n}) = {1 \over {\rm Prob}[{\rm all \ } h_{i} \leq 4]} = 
{1 \over 1 - {\rm Prob}[h_{z_1} = 5 \vee \ldots \vee h_{z_n} = 5]}.
\ee

For $n=1$, and when the boundary condition on the real axis is all open, the
value of $F(z)$ can easily be worked out for $z$ far from the boundary, 
\be
F_{\rm op}(z) = 1 + \Delta_{\rm op}^{-1}(z;z) = {1 \over 2\pi} \log|z-\bar z| 
+ {1 \over 2\pi} \,(\gamma + {3 \over 2} \log{2}) + 1 + \ldots
\label{fullop}
\ee
where the ellipses stand for corrections that go to 0 when $|z-\bar z|$ goes to
infinity (correction terms to scaling), and $\gamma=0.57721\ldots$ is the Euler
constant.

For the all closed boundary condition, the fraction $F_{\rm cl}(z)$
diverges, as do all higher point functions $F_{\rm cl}(z_{1},\ldots,z_{n})$.
Indeed since the sites $z_{i}$ are the only sink sites, the relaxation process
will produce a constant flow of sand towards them, making at least one of them
almost always `full', that is, ${\rm Prob}[h_{z_1} = 5 \vee \ldots \vee h_{z_n} = 5]
= 1$. If one regularizes the fraction like in Section 3, by picking the coefficient
of $2\Delta^{-1}_{\rm plane}(0;0)$ in $F_{\rm cl}(z) = 1 + \Delta_{\rm cl}^{-1}(z;z)$,
then the regularized fraction is equal to~1. 

These two simple calculations are consistent with identifying the introduction
of dissipation at $z$ with the insertion of a logarithmic field $\omega(z,\bar z)$.
As recalled in Section 2, this is also supported by the fact that $\int {\rm 
d}^{2}z \, \omega(z,\bar z)$ is the perturbation term that drives the conformal action to a
massive regime, when dissipation is added at all sites of the lattice. Then, 
\be
\la \omega(z,\bar z) \ra_{\rm op} = {1 \over 2\pi} \log{|z-\bar z|}
+ \gamma_{0}, \qquad \la \omega(z,\bar z) \ra_{\rm cl} = 1,
\label{1pt}
\ee
with $\gamma_{0} = {1 \over 2\pi} \,(\gamma + {3 \over 2} \log{2}) + 1$.

For $n=2$, the calculation of $F(z_{1},z_{2})$ for an all open boundary
yields, in the scaling limit, what should be the 2--point function
\be
\la \omega(z_{1}) \omega(z_{2}) \ra_{\rm op} =
-{1 \over 4\pi^{2}} \log^{2}{|z_{1}-z_{2}| \over |z_{1}-\bar z_{2}|} +
\Big ({1 \over 2\pi} \log{|z_{1}- \bar z_{1}|} + \gamma_{0}\Big) \Big({1 \over
2\pi} \log{|z_{2}- \bar z_{2}|} + \gamma_{0}\Big).
\label{2op}
\ee
When the boundary is all closed, the regularized fraction gives
\be
\la \omega(z_{1}) \omega(z_{2}) \ra_{\rm cl}  = {1 \over \pi} \log{|z_{12}|} 
+ 2\gamma_{0} + {1 \over 2\pi} \log{|z_{1}-\bar z_{2}|^{2} \over |z_{1}-\bar
z_{1}| |z_{2}-\bar z_{2}|}.
\label{2cl}
\ee

These correlators allow to compute the bulk OPE of $\omega$ with itself. Assuming that  
it transforms like $\omega(z,\bar z) \to \omega(w,\bar w) + \lambda
\log{|{{\rm d}w \over {\rm d}z}|^{2}}$ (its normalization is fixed by (\ref{1pt})),
then the M\"obius invariance fixes the general form of the OPE,
\be
\omega(z_{1}) \omega(z_{2}) = \Big[a - 2 \lambda \log{|z_{12}|^{2}}\Big] \,
\omega(z_{2})
+ \Big[b + a \lambda \log{|z_{12}|^{2}} - \lambda^{2}
\log^{2}{|z_{12}|^{2}}\Big] \, \bi(z_{2}) + \ldots
\label{buope}
\ee
for two constants $a,b$ and where the dots stand for terms which vanish when
$z_{1}=z_{2}$. A simple comparison with (\ref{2op}) then yields, using
(\ref{1pt}),
\be
\lambda = -{1 \over 4\pi}, \qquad a = 2\gamma_{0}, \qquad b = -{a^{2} \over 4}
= -\gamma_{0}^{2}.
\ee
Then the expectation value of the OPE in front of a closed boundary exactly reproduces
(\ref{2cl}), up to a term which vanishes if $z_{1}=z_{2}$. This not only provides
a consistent check on the OPE, but also of the regularization prescription we
have used throughout for an all closed boundary.

The problem of identifying the field corresponding to the insertion of
dissipation at a boundary site, and the corresponding boundary OPE, may be
analyzed along the same lines. We consider a closed boundary only, as
sites on an open boundary are already dissipative. 

That field must belong to the representation $\R_{0}$ of the $c=-2$ theory,
and anticipating a little bit, it is not difficult to see that it is a weight
zero logarithmic field, which we call $\omega_{b}$ (hence a logarithmic partner of
the identity). Indeed the regularized fractions $F(x_{1},\ldots,x_{n})$,
for all $x_{i}$ on the boundary, are supposed to converge in the scaling limit to
the $n$--point function $\la \omega_{b}(x_{1}) \ldots \omega_{b}(x_{n})
\ra_{\rm cl}$, and one finds, for $n=1,2,3$, 
\bea
&& \la \omega_{b}(x) \ra_{\rm cl} = 1, \\
&& \la \omega_{b}(x_{1}) \omega_{b}(x_{2}) \ra_{\rm cl} = {2 \over \pi} \,
\log{|x_{12}|} + 4 \gamma_{0} - {5 \over 2},  \label{bb}\\
&& \la \omega_{b}(x_{1}) \omega_{b}(x_{2}) \omega_{b}(x_{3}) \ra_{\rm cl} = 
- \Big({1 \over \pi} \, \log{|x_{12}|} + 2 \gamma_{0} - {5 \over 4} + 
{1 \over \pi}\, \log{|{x_{13} \over x_{23}}|} \Big)^{2} + \nonumber\\
&& \hspace{4.5cm} \Big({2 \over \pi} \, \log{|x_{12}|} + 4 \gamma_{0} -              {5
\over 2}\Big) \Big({2 \over \pi} \, \log{|x_{13}|} + 4 \gamma_{0} - {5 \over
2}\Big).
\eea
They univoquely fix the first terms in the OPE of $\omega_{b}$ with itself,
given on general grounds by a chiral version of (\ref{buope}),
\be
\omega_{b}(x_{1}) \omega_{b}(x_{2}) = \Big[a_{b} - 2 \lambda_{b}
\log{|x_{12}|}\Big] \, \omega_{b}(x_{2}) + \Big[b_{b} + a_{b} \lambda_{b}
\log{|x_{12}|} - \lambda_{b}^{2} \log^{2}{|x_{12}|}\Big] \, \bi_{N}(x_{2}) +
\ldots
\ee

A straightforward comparison with the 1--, 2-- and 3--point functions yields the
value of the three coefficients
\be
\lambda_{b} = -{1 \over \pi}, \qquad a_{b} = 4\gamma_{0} - {5 \over 2}, \qquad
b_{b} = -{a_{b}^{2} \over 4}.
\ee

The equality $\lambda_{b}=\lambda_{N}$ (see Eq. (\ref{lN})) is not a
coincidence. The cylinder Hilbert space with closed boundary condition on both
edges contains a single copy of the representation $\R_{0}$ \cite{r}. Therefore
the two fields $\omega_{b}$ and $\omega_{N}$ must be
(almost) proportional, and since their normalizations are identical, their
conformal transformations must be the same. In fact, we will see in the next 
section that they are not quite identical, but rather differ by a multiple of the
identity $\bi_{N}$, which explains why we gave them different names. On the
other hand, there is only one possible field in $\R_{0}$ which transforms like
the identity $\bi_{N}$, so that there is no ambiguity for the identity term.

Finally one may examine how the bulk dissipation field $\omega(z,\bar z)$ close
to a boundary expands on boundary fields. It must expand on fields of $\V_{0}$ 
close to an open boundary and on fields of $\R_{0}$ close to a closed boundary.
Mo\"ebius invariance again fixes the precise form of the first terms in the
expansion, for $z = x + i y$,
\be
\omega(z,\bar z) = c \, \omega_{b}(x) + \Big[d + (c\lambda_{b} - 2
\lambda)\, \log{2y} \Big] \, \bi(x) + \ldots 
\ee
where $\lambda = -{1 \over 4\pi}$ and $\lambda_{b} = -{1 \over \pi}$. The
identity $\bi(x)$ can either be $\bi_{D}(x))$ or $\bi_{N}(x)$ depending on the
type of boundary. 

If it is an open boundary, the coefficient $c$ is equal to 0, and the 1--point
function (\ref{1pt}) readily gives $d = \gamma_{0}$. In particular, one sees
that the field corresponding to the addition of dissipation at an open site
must be a descendant of the identity $\bi_{D}$. It has been identified in 
\cite{jeng2} as the primary weight 2 field proportional to $\d \t \d \tb$ in the
lagrangian realization (\ref{lag}).

If $\omega(z,\bar z)$ is close to a closed boundary, the simplest way to
determine the coefficients is from the 2--point function (\ref{2cl}) in the
limit where $z_{1}$ and $z_{2}$ approach the boundary. One easily finds $c = 1$
and $d = {5 \over 4} - \gamma_{0}$.

Summarizing, one has
\be
\omega(z,\bar z) = \omega_{b}(x) + \Big[{5 \over 4} - \gamma_{0} - 
{1 \over 2\pi} \, \log{2y} \Big] \, \bi_{N}(x) + \ldots + \Big[\gamma_{0} + {1
\over 2\pi} \log{2y}  \Big] \, \bi_{D}(x) + \ldots
\label{bope}
\ee


\section{Dissipation with change of boundary condition}

In order to probe finer effects of the insertion of isolated dissipation, we
consider, as in Section 3 and 4, the cases of an open boundary with a closed
interval $I=[z_1,z_2]$, and the inverse situation, and the corresponding
fractions $F_{\rm op}(z_1,z_2;z)$ and $F_{\rm cl}(z_1,z_2;z)$. They measure
the effect of inserting a unit of dissipation at site $z$ in presence of two boundary
conditions, and are still given by an expression like (\ref{gen}), where $\Delta
= \Delta_{\rm op} - B_I$ or $\Delta = \Delta_{\rm cl} + B_I$, 
\be
F_{\rm op}(z_1,z_2;z) = {\det{[\bi - \Delta_{\rm op}^{-1}\,(B_I - D_z)]}
\over
\det{[\bi - \Delta_{\rm op}^{-1}\,B_I]}}\,, \quad 
F_{\rm cl}(z_1,z_2;z) = {\det{[\bi + \Delta_{\rm cl}^{-1}\,(B_I + D_z)]} \over
\det{[\bi + \Delta_{\rm cl}^{-1}\,B_I]}}\,.
\label{pmopcl}
\ee

\begin{figure}[h]
\leavevmode
\begin{center}
\mbox{\includegraphics[width=8cm]{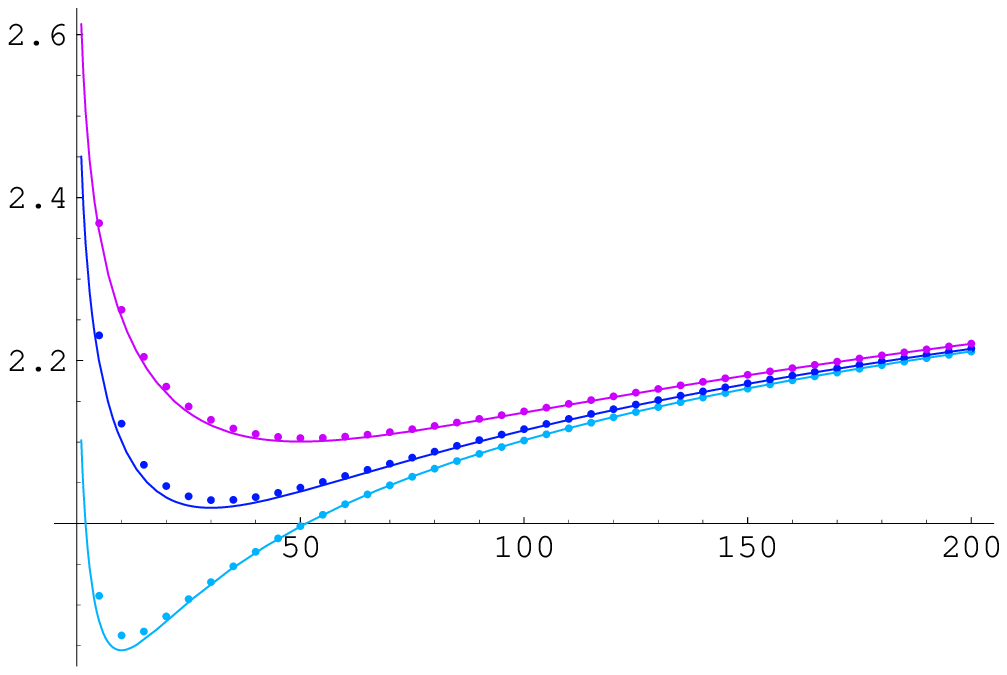}} \hfill
\mbox{\includegraphics[width=8cm]{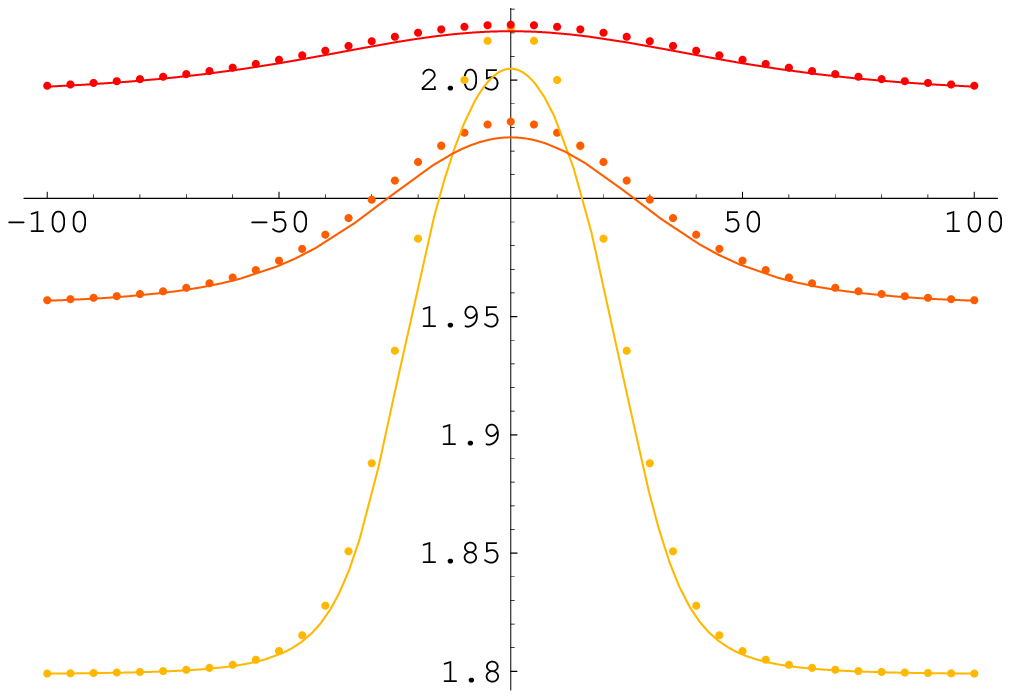}}
\end{center}
\caption{\small Values of $F_{\rm op}(-R,R;x+iy)$ as function of $y$ (left) or as
function of $x$ (right), either computed numerically (dots) or from CFT
computations, Eq. (\ref{fop}) (solid lines). The curves on the left correspond to
$x=0$ and $5 \leq y \leq 200$, for values of $R$ equal to $R=10$ (light blue), $R=30$
(blue) and $R=50$ (violet). Those on the right correspond to $R=30$ and $y=15$
(light orange), $y=40$ (orange) and $y=70$ (red), with $-100 \leq x \leq 100$.}
\end{figure}

The numerical calculation of these two functions is fairly straightforward.
For an interval $I=[-R,R]$ symmetric around the origin, we have
computed the values of $F_{\rm op}(-R,R;z)$ and $F_{\rm cl}(-R,R;z)$
(regularized as before) for $z$ on vertical and horizontal slices in the UHP, and
for different values of $R$ (same as in Section 4). The results are given in
Figure 4 and 5, as colour dots. In both figures, the plots on the left correspond
to the vertical slice $\Re z = 0$ and for three values $R=10, 30, 50$, while the
plots on the right correspond to $R=30$ for the three  horizontal  slices $\Im z
= 15, 40$ and 70.
\begin{figure}[h]
\leavevmode
\begin{center}
\mbox{\includegraphics[width=8cm]{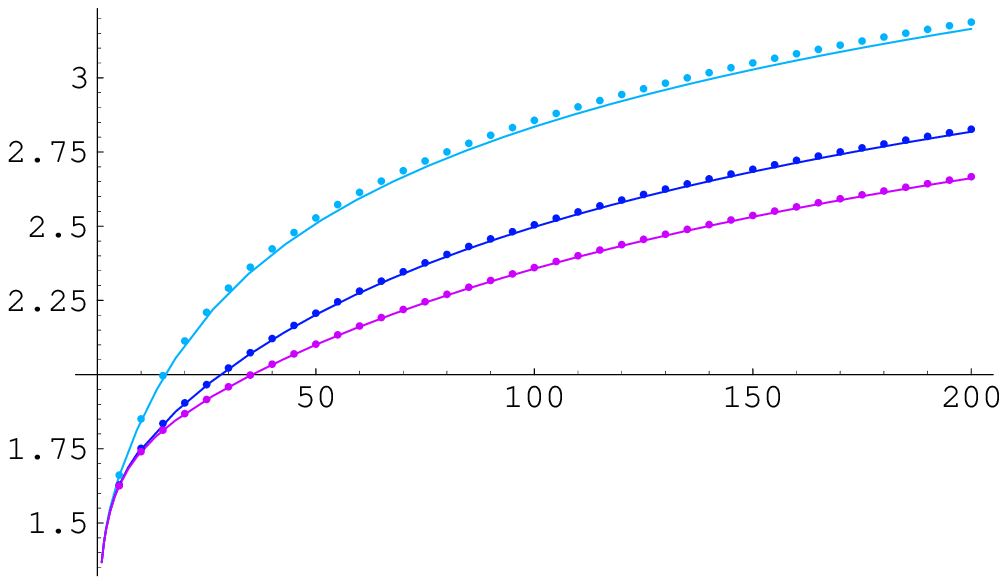}} \hfill
\mbox{\includegraphics[width=8cm]{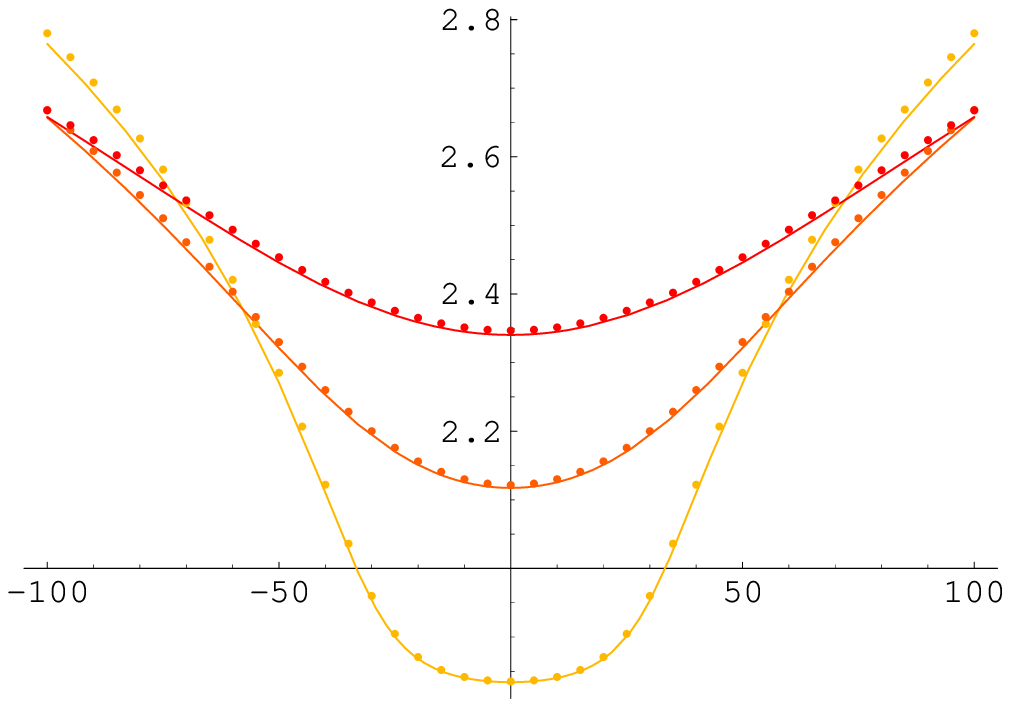}}
\end{center}
\caption{\small Values of $F_{\rm cl}(-R,R;x+iy)$ as function of $y$ (left) or as
function of $x$ (right), numerically (colour dots) or from CFT calculations (solid
lines). The curves on the left correspond to $x=0$ and $5 \leq
y \leq 200$, for values of $R$ equal to $R=10$ (light blue), $R=30$ (blue) and $R=50$
(violet). Those on the right correspond to $R=30$ and $y=15$ (light orange), $y=40$
(orange) and $y=70$ (red), with $-100 \leq x \leq 100$.}
\end{figure}

In the conformal theory, the fractions $F(z_1,z_2;z)$ should correspond to a 3--point
function in the UHP. The two jumps of boundary condition on the real axis are
effected by the insertion of two $h=-{1 \over 8}$ primary fields, namely one
$\mdn$ and one $\mnd$, whereas the insertion of dissipation is represented by
the insertion of the logarithmic field $\omega(z,\bar z)$ (see previous section).

Then the scaling limits of the lattice fractions are given by 3--point
correlators on the UHP or 4--point chiral correlators on the plane 
\be
F(z_1,z_2;z) = {1 \over A z_{12}^{1/4}} \, \la \mu(z_1) \,
\mu(z_2) \, \omega(z) \, \omega(z^{*})\ra_{\rm chiral,plane}\,,
\ee
where the $\mu$ fields are chosen according to the case we want to consider. We
first compute the general form of the 4--point function, and then choose the
particular solution which suits the boundary conditions we impose on the real
axis.

Because $\omega$ is the logarithmic partner of a primary field (the identity), the
4--point function requires a multiple step calculation. Indeed it satisfies an
inhomogeneous differential equation (see for instance \cite{flohr}), where the
inhomogeneity depends on the correlators where each logarithmic field $\omega$ is
in turn replaced by its primary partner $\lambda \bi$, with $\lambda = -{1 \over
4\pi}$. One finds from the 3--point function
\be
\la \mu(z_1) \, \mu(z_2) \, \omega(z_3) \ra = z_{12}^{1/4} \, \Big[
\beta + \alpha \lambda \, \log{z_{12} \over z_{13}z_{23}}\Big],
\ee
the general form of the chiral 4--point correlator, with the same cross ratio $x
= {z_{12} z_{34} \over z_{13} z_{24}}$ as before, 
\bea
\la \mu(z_1) \, \mu(z_2) \, \omega(z_3) \, \omega(z_4)\ra  &=& z_{12}^{1/4} \,
\Big\{ \gamma - \delta \, \log{1 - \sqrt{1-x} \over 1+ \sqrt{1-x}}
- \alpha \lambda^{2}\,\Big(\log{1 - \sqrt{1-x} \over 1+ \sqrt{1-x}}\Big)^2
\nonumber\\
&+& \beta \lambda \, \log{x^2 \over 1-x} - 2\beta \lambda \, \log{z_{34}}
+ \alpha \lambda^{2} \, \log{z_{21} \over z_{13} z_{23}} \, \log{z_{21} \over
z_{14} z_{24}}\Big\}\,,
\label{mmoo}
\eea
where $\alpha,\beta,\gamma,\delta$ are arbitrary constants. 

For the case ``closed interval in an open boundary'', the function 
\be
F_{\rm op}(z_1,z_2;z) = {1 \over A  z_{12}^{1/4}} \, \la \mdn(z_1) \,
\mnd(z_2) \, \omega(z) \, \omega(z^{*})\ra\,,
\label{fgop}
\ee
can be determined by choosing the solution (\ref{mmoo}) which has no logarithmic
singularity when $z_{1},z_{2}$ go to 0, and which reproduces the fraction $F_{\rm 
op}(z)$ given in (\ref{fullop}).

The first condition forces $\alpha=0$ and $\delta=2\beta \lambda$ (remember that
$\sqrt{1-x}$ goes to $+1$), whereas the second one imposes 
\be
\beta = A \,, \qquad \gamma = A\,(\gamma_{0} + {1 \over \pi} \log{2})\,.
\ee

This provides a completely explicit expression for $F_{\rm op}(z_1,z_2;z)$,
\be
F_{\rm op}(z_1,z_2;z) = \gamma_{0} + {1 \over \pi} \log{2} -  
{1 \over 2\pi} \, \log{(1 +
\sqrt{1-x})^2 \over \sqrt{1-x}} + {1 \over 2\pi} \, \log{|z - z^{*}|}\,.
\ee
For $z_1=-R$ and $z_2=R$, and for $z=x+iy$, it reduces to
\be
F_{\rm op}(-R,R;z) = 1 + {1 \over 2\pi} \,(\gamma + {7 \over 2} \log{2}) 
- {1 \over 2\pi} \log\Big\{{1-{R^2-|z|^2 \over |R^2-z^2|} \over y} \Big\}\,.
\label{fop}
\ee

For the cases worked out numerically and discussed above, the previous formula
yields the solid curves shown in Figure 4.

The expansion of $F_{\rm op}(-R,R;z)$ in the two regimes $R$ small and $R$ large
read
\be
F_{\rm op}(-R,R;z) = \cases{
1 + {1 \over 2\pi} \,(\gamma + {5 \over 2} \log{2}) + {1 \over 2\pi} \, \log{y} +
\ldots & for $R \ll 1$, \cr
1 + {1 \over 2\pi} \,(\gamma + {5 \over 2} \log{2} + \log{R^2}) - {1 \over 2\pi}
\, \log{y} + \ldots & for $R \gg 1$.}
\ee
The large $R$ limit corresponds to shrinking the open portion to nothing, and
therefore to fusing the two $\mu$ fields at infinity. From (\ref{fusN}), the
fusion gives rise to two channels, proportional to $\log{R}$ and to~1.

The fraction for the opposite case, that of a closed boundary containing an open interval,
\be
F_{\rm cl}(z_1,z_2;z) = {1 \over A z_{12}^{1/4}} \, \la \mnd(z_1) \,
\mdn(z_2) \, \omega(z) \, \omega(z^{*})\ra\,,
\label{fcl}
\ee
is related to the previous one by the inversion $z \to -1/z$. Since the correlators
are computed so as to be invariant under the (inhomogeneous) Mo\"ebius
transformations, the only change comes from the monodromy properties discussed
in Section 4, which simply change the sign of $\sqrt{1-x}$. One thus obtains
\be
F_{\rm cl}(z_1,z_2;z) = \gamma_{0} + {1 \over \pi} \log{2} - {1 \over 2\pi} \, 
\log{\Big|{(1 - \sqrt{1-x})^2 \over \sqrt{1-x}}\Big|} + {1 \over 2\pi} \, \log{|z -
z^{*}|}\,.
\ee
Setting $z_1=-R$ and $z_2=R$, we obtain the formula,
\be
F_{\rm cl}(-R,R;z) = 1 + {1 \over 2\pi} \,(\gamma + {7 \over 2} \log{2}) 
- {1 \over 2\pi} \log\Big\{{1+{R^2-|z|^2 \over |R^2-z^2|} \over y} \Big\}\,,
\ee
which has been used to generate the solid lines pictured in Figure 5. The agreement
is excellent. The expansion of $F_{\rm cl}(-R,R;z)$ in the two extreme regimes now read
\be
F_{\rm cl}(-R,R;z) = \cases{
1 + {1 \over 2\pi} \,(\gamma + {5 \over 2} \log{2} - \log{R^2}) - {1 \over 2\pi}
\, \log{y \over |z|^4} + \ldots & for $R \ll 1$, \cr
1 + {1 \over 2\pi} \,(\gamma + {5 \over 2} \log{2}) + {1 \over 2\pi}
\, \log{y} + \ldots & for $R \gg 1$.}
\ee

We finish this section with two comments, and first on the $R \ll 1$ limit in
the previous expression. 

According to the fusion of two $\mu$ fields on a closed boundary, the
logarithmic term $-{1 \over \pi} \log{R}$ should be proportional to
$\lambda_{N} \la \omega(z,\bar z) \ra_{\rm cl} \, \log{R}$ and confirms all
previous results. On the other hand, the non--logarithmic piece should
correspond to $\la \omega_{N}(0) \, \omega(z,\bar z) \ra_{\rm cl}$, namely a
chiral 3--point function on the plane $\la \omega_{N}(0) \, \omega(z) \omega(\bar
z) \ra$. Its $z$ dependence is however unusual and is due to the fact that
the two logarithmic fields involved have different inhomogeneous terms in
their conformal transformations, $\lambda_{N} \neq \lambda$. An explicit
calculation shows that the 3--point function in this situation is exactly what
the above limit yields. 

Our second remark is on the relation between the two logarithmic boundary
fields, $\omega_{N}$ and $\omega_{b}$. Both have the same normalization $\la
\omega_{b}(x)\ra = \la \omega_{N}(x) \ra = 1$, and the same conformal
transformations $\lambda_{N} = \lambda_{b} = -{1 \over \pi}$, yet they do not
have the same 2--point function. From the limit $z_{12}, z_{34} \to 0$ of the
four $\mu$ correlator (\ref{nd}), one obtains
\be
\la \omega_{N}(x_{1}) \omega_{N}(x_{2}) \ra_{\rm cl} = {2\over \pi}
\log{|x_{12}|} + {4 \over \pi} \log{2},
\ee
which differs from $\la \omega_{b}(x_{1}) \omega_{b}(x_{2}) \ra$ in (\ref{bb})
by the constant piece. The two fields belonging to the same representation,
they can only differ by a multiple of the identity. Comparing the two 2--point
functions, one finds
\be
\omega_{b} = \omega_{N} + \kappa \, \bi_{N}\,, \qquad \kappa = 
{3 \over 4} + {1 \over \pi} (\gamma - {1 \over 2}\log{2}).
\ee
One obtains the same value of $\kappa$ from the mixed 2--point function arising
in the limit $z_{12} \to 0$, $z - z^{*} \to 0$ of the correlator (\ref{fcl}), when using
the field decomposition (\ref{bope}).


\section{Conclusions}

Our primary motivation in this article was to contribute to establishing the Abelian
sandpile model as a model that can be described by a conformal field theory. Whether
and to what extent the conformal description holds remains an unclear issue, mainly 
because of the highly non--local interactions present in the sandpile model. 

We have essentially considered three operators. The first is the operator that changes
a boundary condition between open and closed; the second corresponds to the height 1 
variable; and the third one is the insertion of dissipation at a non--dissipative site. We
have studied fine details of their mixed correlations on the upper half plane, and found
indeed a full agreement with the predictions of a logarithmic conformal theory with 
central charge $c=-2$. 

Though our results are certainly encouraging and show that the conformal description
cannot be all wrong, it is still however far from proving that every aspect of the sandpile
model has a counterpart in the conformal theory and vice--versa. Before this goal may be
attained, important questions must be answered, like the existence of other
boundary conditions than open and closed, the sandpile interpretation of the weight 
$3/8$ primary field, the sandpile significance of the $Z_{2}$ symmetry and of the 
$\cal W$--algebra present in the conformal theory, the conformal description of the higher height
variables and of the avalanche observables. The list is not exhaustive but only shows the
benefit one can expect on both sides.


\appendix
\section*{Appendix}
\setcounter{section}{1}

We give here some details on the numerical calculations reported in the text. All
numerical results are related to the calculation of finite determinants, the largest
ones being of a typical size of 100, though larger ones have been considered.

The matrices of which the determinant is to be computed are of the form $(\bi +
\Delta_{\rm op}^{-1}A)$ or $(\bi + \Delta_{\rm cl}^{-1}A)$ where $A$ is a numerical 
matrix. Their entries are linear combinations of $(\Delta_{\rm op})^{-1}_{ij}$
and $(\Delta_{\rm cl})^{-1}_{ij}$, where $i,j \in \Z \times \Z_{>}$ run over some set
of lattice sites in the upper half--plane, the boundary of which is the horizontal line
$y=1$. By the method of images, the inverse matrices $\Delta_{\rm op}^{-1}$ and 
$\Delta_{\rm cl}^{-1}$ are related to the inverse Laplacian on the full plane
$\Delta_{\rm plane}^{-1}$. If $(m_{1},n_{1})$ and $(m_{2},n_{2})$ are the integer
coordinates of $i$ and $j$ respectively, then
\bea
(\Delta_{\rm op})^{-1}_{(m_{1},n_{1}),(m_{2},n_{2})} &=& 
(\Delta _{\rm plane})^{-1}_{(m_{1},n_{1}),(m_{2},n_{2})} - 
(\Delta _{\rm plane})^{-1}_{(m_{1},n_{1}),(m_{2},-n_{2})}, \\
(\Delta _{\rm cl})^{-1}_{(m_{1},n_{1}),(m_{2},n_{2})} &=& 
(\Delta _{\rm plane})^{-1}_{(m_{1},n_{1}),(m_{2},n_{2})} + 
(\Delta _{\rm plane})^{-1}_{(m_{1},n_{1}),(m_{2},1-n_{2})}.
\eea

Thus the entries of the inverse Laplacian on the plane $\Z^{2}$ are required. Because of
the horizontal and vertical symmetries, it is enough to know $(\Delta_{\rm
plane})^{-1}_{ij}$ where one site is the origin. A simple Fourier transformation yields a
divergent integral representation
\be
(\Delta_{\rm plane})^{-1}_{(m,n),(0,0)} = \int\!\!\!\!\int_0^{2\pi} \; 
{{\rm d}^2k \over 4\pi^2} \; {{\rm e}^{ik_1 m + ik_2 n} \over 4 - 2\cos{k_1} - 
2\cos{k_2}},
\ee
and a finite representation for the difference 
\be
\phi(m,n) \equiv (\Delta_{\rm plane})^{-1}_{(m,n),(0,0)} - 
(\Delta_{\rm plane})^{-1}_{(0,0),(0,0)}
= \int\!\!\!\!\int_0^{2\pi} \; {{\rm d}^2k \over 4\pi^2} \; 
{{\rm e}^{ik_1 m + ik_2 n} - 1 \over 4 - 2\cos{k_1} -  2\cos{k_2}}.
\ee
So the entries of $\Delta_{\rm op}^{-1}$ are well--defined and finite. On the other
hand,  the entries $\Delta_{\rm cl}^{-1}$, being a sum of two
inverse Laplacian entries, have all an infinite piece, which can be taken as 
$2(\Delta_{\rm plane})^{-1}_{(0,0),(0,0)}$. The regularized determinant (\ref{reg}) is
however finite as its entries are differences of $(\Delta_{\rm cl})^{-1}_{ij}$. 

>From the reflection symmetries, $\phi(m,n) = \phi(-m,n) = \phi(m,-n) = \phi(n,m)$, it is
enough to known $\phi(m,n)$ in the first half--quadrant delimited by the positive
horizontal axis $m \geq 0, n=0$, and the diagonal $m=n \geq 0$. A convenient procedure 
\cite{spitz} to compute $\phi(m,n)$ in that region is to use the known exact values on the
diagonal, given by
\be
\phi(m,m) = -{1 \over \pi} \sum_{k=1}^{m} {1 \over 2k-1}, \qquad (m \geq 1)
\ee
and then to propagate the function $\phi$ from the diagonal down to the $x$--axis by a
repeated use of the Poisson equation, 
\be
4\phi(m,n) - \phi(m+1,n) - \phi(m-1,n) - \phi(m,n+1) - \phi(m,n-1) = 
\delta_{m,0} \, \delta_{n,0}.
\ee
In this way the knowledge of $\phi(m,n)$ for fixed $m$ and for $0 \leq n \leq m$, and of
the diagonal element $\phi(m+1,m+1)$, allows to determine all the values of $\phi$ on the
next line, namely $\phi(m+1,n)$ for $0 \leq n \leq m+1$. 

This way of propagating the function $\phi$ is numerically unstable because the
Poisson equation involves the difference of close numbers. If the propagation is not
performed with enough numerical precision, the resulting values of $\phi$ depart very
wildly from what is expected. In the computations reported in the text, the values of
$\phi(m,n)$ are required for values of $m$ of the order of 400. For the above propagating
procedure to produce sensible results, all calculations were performed on 320 decimal
places. 

Once the actual values of $\phi(m,n)$ are obtained, the determinants can be computed.
All determinants considered in the text diverge exponentially, or go to 0 exponentially,
with their
size. Consider for instance $\det[\bi - \Delta_{\rm op}^{-1}B_{I}]$, where $I$ is an
interval on the boundary, possibly disconnected (like in (\ref{ellop})). In the sandpile
model, it is equal to the number of recurrent configurations when the boundary is
open except on the segment $I$ which is closed, divided by the corresponding number with
an all open boundary. Because a closed boundary site has a free energy smaller
than an open boundary site, by an amount equal to $2G/\pi$ \cite{r}, the determinant is
dominated by an exponentially small term $e^{-2G |I|/\pi}$ ($G$ is the Catalan
constant). For the same reason, the determinant $\det[\bi + \Delta_{\rm cl}^{-1}B_{I}]$
for the converse situation is dominated by an exponentially diverging term $e^{2G
|I|/\pi}$. The same is true if a matrix $B_{z}$ (relevant for a unit height) or $D_{z}$
(used for an isolated dissipative site) is added to $B_{I}$.

These exponential factors drop out in ratios of partition functions eventually related to
4--point CFT correlators ---namely the ratios in Eqs. (\ref{ellop}), (\ref{ellcl}),
(\ref{p1opcl}) and (\ref{pmopcl})---, but taking the ratio of huge numbers is not
numerically efficient. To avoid this problem, one multiplies the matrices by the proper
factor $e^{\pm 2G/\pi}$ before computing its determinant, so as to kill the dominant
exponentials. 

As the determinant calculations generate a moderate loss of precision, the precision on
the matrix entries is at this stage lowered to 25 decimal places. The numerical errors
on the final results are expected to be smaller than 0.001\%. In view of the relative
importance of the corrections to scaling, there is no need to improve it.


\section*{References}
\pdfbookmark[1]{References}{table}

\end{document}